\documentclass[%
 aip,
 amsmath,amssymb,
 reprint,%
]{revtex4-1}

\usepackage{graphicx}
\usepackage{dcolumn}
\usepackage{bm}

\usepackage[utf8]{inputenc}
\usepackage[T1]{fontenc}
\usepackage{mathptmx}
\usepackage{etoolbox}

\makeatletter
\def\@email#1#2{%
 \endgroup
 \patchcmd{\titleblock@produce}
  {\frontmatter@RRAPformat}
  {\frontmatter@RRAPformat{\produce@RRAP{*#1\href{mailto:#2}{#2}}}\frontmatter@RRAPformat}
  {}{}
}%

\preprint{AIP/123-QED}

\usepackage{graphicx}
\usepackage{dcolumn}
\usepackage{bm}
\usepackage{subfigure}
\usepackage{multirow}
\usepackage{soul}
\usepackage[normalem]{ulem}
\usepackage{xcolor}
\usepackage{kantlipsum}
\usepackage{textcomp}

\usepackage{booktabs}
\newcolumntype{C}{>{$}c<{$}}
\AtBeginDocument{
\heavyrulewidth=.08em
\lightrulewidth=.05em
\cmidrulewidth=.03em
\belowrulesep=.65ex
\belowbottomsep=0pt
\aboverulesep=.4ex
\abovetopsep=0pt
\cmidrulesep=\doublerulesep
\cmidrulekern=.5em
\defaultaddspace=.5em
}
\usepackage{siunitx}
\usepackage{multirow}

\definecolor{amber}{rgb}{1,0.49,0}

\newcommand{\editor}[2]{%
  \expandafter\newcommand\csname #1note\endcsname[1]{%
    \textcolor{#2}{(\textbf{#1:} ##1)}}%
  \expandafter\newcommand\csname #1\endcsname[1]{%
    \textcolor{#2}{##1}}%
  \expandafter\newcommand\csname #1cancel\endcsname[1]{%
    \textcolor{#2}{\sout{##1}}}%
  \expandafter\newcommand\csname #1change\endcsname[2]{%
    \textcolor{#2}{\sout{##1} ##2}}%
  \newenvironment{#1text}{\color{#2}}{\color{black}}
}

\def\eqref#1{(\ref{#1})}

\def\PbF2{{PbF$_2$}}
\def\CaF2{{CaF$_2$}}
\def\UO2{{UO$_2$}}
\def\lammps{{\textsc{lammps}}}

\editor{FG}{blue}
\editor{resub}{cyan}

\usepackage{mathrsfs}
\usepackage{placeins}

\begin{document}

\title{Investigating finite-size effects in {molecular dynamics} simulations of {ion diffusion, heat transport, and thermal motion in} superionic materials}
\author{Federico Grasselli}
\affiliation{
  COSMO -- Laboratory of Computational Science and Modelling, IMX, \'Ecole Polytechnique F\'ed\'erale de Lausanne, 1015 Lausanne, Switzerland
}
\email{federico.grasselli@epfl.ch}

\date{\today}


\begin{abstract}
The effects of the finite size of the simulation box in equilibrium molecular dynamics simulations are investigated for prototypical superionic conductors of different types, namely the fluorite-structure materials \PbF2, \CaF2, and \UO2 (type II), and the $\alpha$ phase of AgI (type I). Largely validated empirical force-fields are employed to run ns-long simulations and extract general trends for several properties, at increasing size and in a wide temperature range.
This work shows that, for the considered type-II superionic conductors, the diffusivity dramatically depends on the system size and that the superionic regime is shifted to larger temperatures in smaller cells. Furthermore, only simulations of several hundred atoms are able to capture the experimentally-observed, characteristic change in the activation energy of the diffusion process, occurring at the order-disorder transition to the superionic regime. Finite-size effects on ion diffusion are instead much weaker in $\alpha$-AgI. 
The thermal conductivity is found generally smaller for smaller cells, where the temperature-independent (Allen-Feldman) regime is also reached at significantly lower temperatures.
The finite-size effects on the thermal motion of the non-mobile ions composing the solid matrix follow the simple law which holds for solids.
\end{abstract}

\pacs{}

\maketitle

\section{Introduction}
\label{sec:intro}

Superionic (SI) materials are characterized by a matrix of atoms arranged in a  (crystalline or amorphous) solid, and of one (or more) mobile species, which diffuses above some critical temperature. 
The interest for SI materials has largely increased during the last decades, along with the quest for good candidates in the realization of solid-state batteries, where charge carriers like lithium ions move through a solid-state electrolytic matrix \cite{Goodenough2010,Janek2016,Kato2016,Kwade2018}. Moreover, the superionic phases of water and ammonia \cite{Cavazzoni1999} have been predicted to compose a large fraction of the outer cores of ice giant planets, like Uranus and Neptune, \cite{Redmer2011,Nettelmann2016} and many recent theoretical \cite{French2011,Grasselli2020} and experimental \cite{Millot2018,Millot2019} studies have focused on transport properties of materials becoming SI at planetary conditions, to study the evolution of these celestial bodies \cite{Podolak2019,Stixrude2021}.

The complexity and the variety of new SI materials naturally imply that, from the computational material-science standpoint, a large effort is devoted to the prediction of static and dynamical properties and their microscopic description by means of atomistic simulations. In particular, molecular dynamics (MD) simulations are needed whenever \textit{dynamical} properties (like transport coefficients and correlation functions) are investigated \cite{Allen2017}. 

Due to the chemical complexity of many of these materials, \textit{ab initio} MD simulations are often performed, whose computational cost currently limits the simulation box to a few hundred atoms at most. This limitation poses serious questions on the role of finite-size effects (FSE) in the characterization of the physical properties of a SI material. For instance, recent tests were run on Li$_{10}$GeP$_2$S$_{12}$-type SI conductors \cite{Huang2021}, where machine-learning interatomic potentials trained on \textit{ab initio} calculations allowed for simulations that are inaccessible to \textit{ab initio} MD \cite{Musil2021}. These calculations showed that simulation boxes containing even some hundred atoms overestimate the Li-ion diffusivity by one order of magnitude with respect to the largest size considered (1600 atoms).
This may have dramatic consequences in calculations aiming to find the best SI conductors for realistic devices \cite{Muy2019,Kahle2020,Materzanini2021}.

In liquids, FSE affecting \textit{particle diffusion} have been extensively investigated (see, e.g., the recent review of Ref.~\onlinecite{Celebi2021}): the hydrodynamics arguments by Yeh and Hummer \cite{Dunweg1993,Yeh2004} suggest that,
for a cubic simulation box in periodic boundary conditions (PBC) and for a given particle density, the diffusivity of the liquid can be corrected by a factor proportional to the inverse of the box side, 
and that the proportionality coefficient only depends on geometric factors, on the temperature, and on the viscosity of the liquid, which is usually largely independent of the size \cite{Yeh2004,Moultos2016}.
Recent works evidenced that the application of the Yeh and Hummer correction is justified also for multicomponent liquids and ionic melts \cite{Jamali2018,Jamali2020,Shao2020}. 
Nevertheless, the hydrodynamics equations of a SI material are different from (and more complicated than) those of simple liquids \cite{Zeyher1978,Dieterich1980}. For instance, transverse modes of the lattice survive, the static shear modulus is non vanishing, and the atoms of the mobile species diffuse via hopping mechanisms that are qualitatively different from the motion of particles in a simple fluid \cite{Dixon1980}: the Yeh-Hummer arguments are therefore inappropriate 
to account for FSE in the diffusion of charge carriers in SI materials. 
{Furthermore, while general trends for the FSE on \textit{heat transport} in solids and liquids have been reported in the literature \footnote{Notice that, even in the case of liquids, the specific functional dependence of the thermal conductivity on the system size seems to be qualitatively affected by the particular pressure and temperature conditions of the simulation \cite{Puligheddu2020}}, such an analysis is currently missing for equilibrium MD simulations of thermal conduction in SI materials}

This article aims at investigating of the FSE in the calculation of relevant static and dynamical properties of SI materials via equilibrium MD simulations. 
I focus on simple yet paradigmatic examples of type-I and type-II SI conductors, which can be effectively described in terms of largely validated empirical potentials. I restrict my analysis to systems with perfect stoichiometry. Furthermore, I consider cubic simulation boxes where the unit cell, of lattice parameter $a$, is equally replicated $\ell$ times in all the three spatial directions, to avoid additional and non-trivial effects which arise, even for simple liquids, in the case of anisotropic replications \cite{Botan2015}. 

In Sec.~\ref{sec:discuss} I discuss the SI materials I selected to investigate. {In Sec.~\ref{sec:methodology} I give methodological details on the equilibrium MD simulations that I performed.}
In Sec.~\ref{sec:results} I provide the main results of the calculations, by analyzing the size dependence of the specific heat capacity, the mobile-ion diffusivity, the thermal conductivity and the Debye-Waller $B$-factors, for each of the selected SI materials.
Finally, I draw general conclusions in  Sec.~\ref{sec:conclusions}.

\section{Discussion}
\label{sec:discuss}

\begin{figure}
    \centering
    \includegraphics[width=0.95\linewidth]{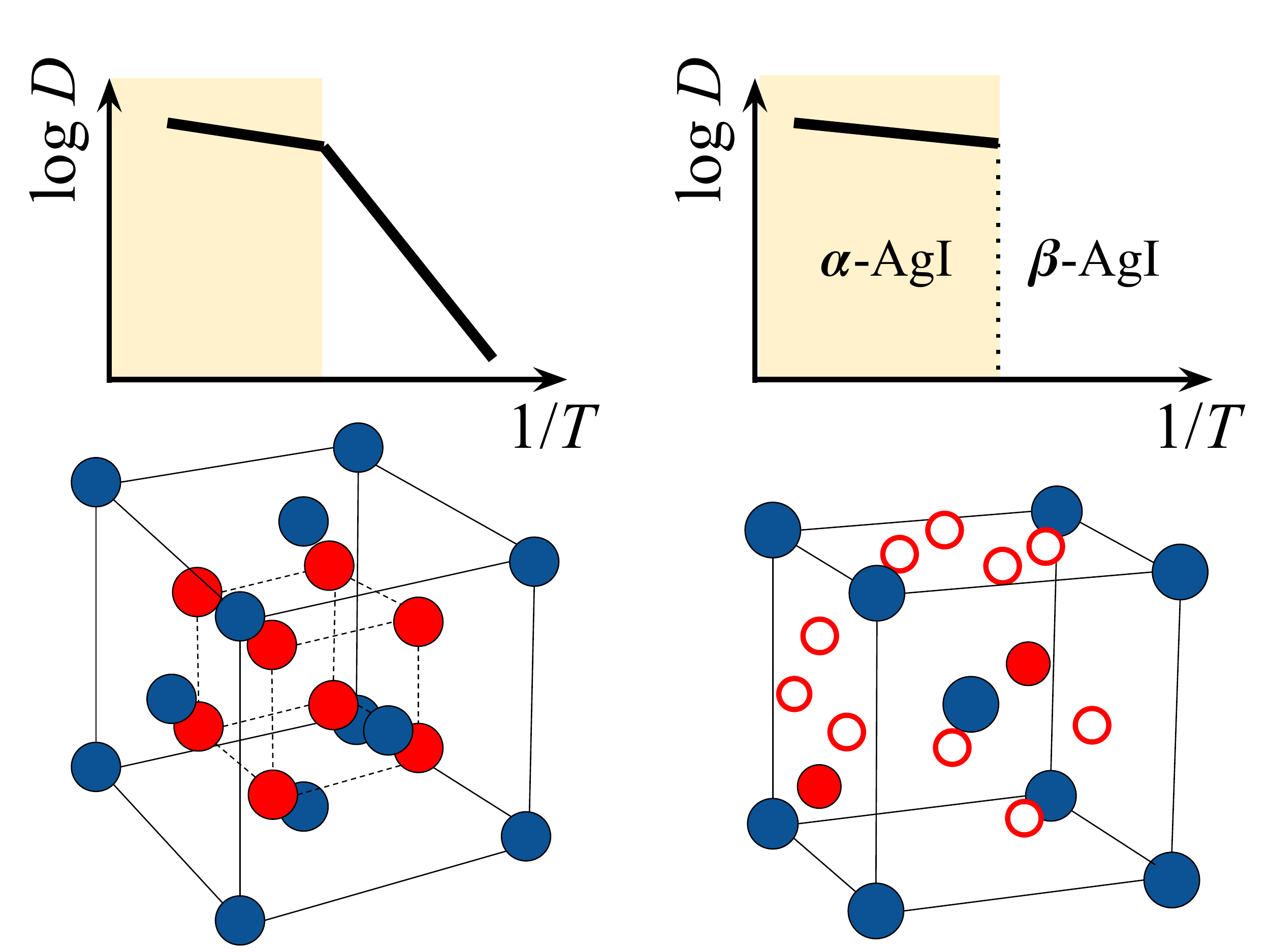}
    \caption{Structure and behavior of the diffusivity, $D$, of the mobile species against inverse temperature for fluorite-structure materials (left) and $\alpha$-AgI (right). The (non-)mobile ions are represented in (blue) red. The empty red circles indicate degenerate tetrahedral positions. Rear-faces' atoms are not displayed. The shaded yellow area indicates the superionic regime. }
    \label{fig:fluorites_AgI}
\end{figure}

I choose the fluorite-structure materials \PbF2, \CaF2, and \UO2, as simple, yet prototypical examples of SI materials where FSE should be particularly relevant: 
all the energetically-equivalent regular sites of the mobile species--the anions--are occupied, and the hopping of one diffusing anion can eventually occur only with a net hopping of other anions, since anion diffusion \textit{``occurs by discrete hops between regular sites''}, and anions \textit{``do not reside in a well-defined manner on the cube-centre sites''} (\textit{verbatim} from Ref.~\onlinecite{Dixon1980}. See also Refs.~\onlinecite{Dixon1980b} for an insightful analysis of anion distribution in fluorites, and Ref.~\onlinecite{Mohn2021} for a recent, comprehensive study of cooperative F dynamics in \PbF2). 
Such a concerted hopping mechanism may easily extend to more than one lattice constant, leading to a size dependence. 
Furthermore, these materials (see Fig.~\ref{fig:fluorites_AgI}, left panels) are characterized by a continuous order-disorder transition to the SI phase with no structural change in the crystalline structure of the non-diffusive species (type II superionic materials). 
The diffusion mechanism depends on the specific temperature regime, and, in particular, whether the system is in the SI phase or not \footnote{Below the SI transition, the transient hopping mechanism is dominated by vacancy motion, while, in the SI phase, it can be \textit{``attributed in roughly equal measure to vacancy and interstitial motion''} (\textit{verbatim} from Ref.~\onlinecite{Gillan1980}).}.

The finite size of the sample is known to cause a shift in the critical temperature of second-order phase transitions \cite{Binder1987}. 
Therefore, for a given temperature and particle density, a small simulation box may be in a different thermodynamic phase with respect to a larger one, with different diffusive mechanisms and a dramatic effect on diffusion. This is likely to be the case in all those materials where the diffusion mechanisms are strongly dependent on $T$. As we shall see in Sec.~\ref{sec:results}, this tangling between the diffusion mechanisms (hydrodynamics) and the phase of the system (thermodynamics) is responsible for dramatic FSE on ionic transport in these materials.

I also investigate a different material, the cubic phase of silver iodide ($\alpha$-AgI), as a typical example of a system where FSE on the diffusion coefficient should be less relevant (Fig.~\ref{fig:fluorites_AgI}, right panels): in contrast with fluorite-structure materials, in $\alpha$-AgI the large degeneracy of equivalent positions that one Ag ion---the diffusive species---may take inside a unit cell results in a large freedom in the choice of empty sites (empty red circles in Fig.~\ref{fig:fluorites_AgI}) that a selected Ag ion can hop to: the temperature affects the probability that hopping occurs, but not the general mechanism of diffusion. Moreover, due to the large degeneracy of empty regular sites, there is no need for the hopping of one Ag cation to be accompanied by the hopping of other neighbor Ag cations.
The $\alpha$-AgI phase is superionic, and is reached after a first-order phase transition (at $\approx 420$ K at ambient pressure\cite{Binner2006}) from the hexagonal, non conducting $\beta$-phase. The sudden, discontinuous change in the ionic diffusion at the phase transition makes AgI a type-I SI conductor.

\section{Methodology}
\label{sec:methodology}

\subsection{Empirical interatomic potentials}
\label{sec:potentials}

The choice of these materials is also motivated by the availability, in the scientific literature, of largely-validated empirical potentials that proved to qualitatively describe the ion diffusion mechanisms as well as the static properties of these systems \cite{Fossati2021}.
These potentials make it possible to run, at a feasible computational cost, reliably long simulations ($\sim$ns) at different temperatures and sizes, to extract general behaviors.  

For \PbF2 and \CaF2, I employ the following two-body potential, combination of Coulomb and Buckingham potentials:
\begin{equation}
    V_{ij} = \frac{z_i z_j e^2}{r} + A_{ij} e^{-r/\rho_{ij}} - \frac{C_{ij}}{r^6} 
    \label{eq:PbF2_CaF2}
\end{equation}
with the parameters optimized in Ref.~\onlinecite{Walker1982} (\PbF2) and in Ref.~\onlinecite{Dixon1980} (\CaF2), reported in Table \ref{tab:PbF2_CaF2}.
\begin{table}[]
    \centering
    \begin{tabular}{
    c
    S[table-format=9.1(1)]
    S[table-format=6.3(1)]
    S[table-format=5.2(0)]
    }
              &   $A_{ij}[\text{eV}]$   &    $\rho_{ij}[\text{\AA}]$   &    
              $C_{ij}[\text{eV} \text{\AA}^{6}]$  \\
    [0.5ex]
        \toprule
        Pb-F  &    122.7 &   0.516  &     0.0  \\
        F-F   &  10225.0 &   0.225  &   107.3  \\
        \midrule
        Ca-F  &   674.3  &  0.336   &   0.0 \\
        F-F   &  1808.0  &  0.293   & 109.1 \\
        \bottomrule
    \end{tabular}
    \caption{Parameters employed in the potential of Eq.~\eqref{eq:PbF2_CaF2}. For cation-cation interaction, only the Coulomb interaction is considered (i.e. $A_{++} = C_{++} = 0$, with $+$ indicating Pb or Ca). The integer charges $z_\mathrm{Pb} = z_\mathrm{Ca} = +2$ and $z_\mathrm{F} = -1$ are used.}
    \label{tab:PbF2_CaF2}
\end{table}
The success of this potential in the microscopic study of ionic diffusion in these materials dates back to the '80s.
For \UO2, I employ a recently developed potential, described in Refs.~\onlinecite{Cooper2014a,Cooper2014b}, which combines Buckingham, Coulomb and Morse potentials to treat two-body interactions, as well as the embedded atom method (EAM) to account for many-body interactions. I refer the interested reader to the original literature and to the Materials Cloud repository of the present work for the explicit parametrization values (see ``Data Availability'').

Finally, for AgI, I use the following combination of Coulomb and Morse potentials:
\begin{equation}
    V_{ij} = \frac{z_i z_j e^2}{r} + D_{ij} \left[ e^{-2\alpha_{ij} (r - R_{ij})} - 2e^{-\alpha_{ij} (r - R_{ij})} \right] 
    \label{eq:AgI}
\end{equation}
with the parameters of Ref.~\cite{Niu2018}, reported in Table \ref{tab:AgI}, derived via the Chen-M\"obius lattice inversion method from \textit{ab initio} calculations of cohesive energies. This potential displays good agreement with experiments \cite{Kvist1970} concerning static properties of different phases, as well as Ag diffusivity in $\alpha$-AgI, and it is also consistent, in a wide temperature range, with the widely-used Parrinello-Vashishta-Rahman empirical potential \cite{Parrinello1983}.   
\begin{table}[]
    \centering
    \begin{tabular}{
    c
    S[table-format=7.2(1)]
    S[table-format=6.3(1)]
    S[table-format=4.2(0)]
    }
             & \textit{D}$_{ij}[\text{eV}]$  & $\alpha_{ij}[\text{\AA}^{-1}]$ & $R_{ij}[\text{\AA}]$ \\
    [0.5ex]
    \toprule
    I-Ag     &  0.55   & 1.600   &  2.6   \\
    I-I      &  0.16   & 0.684   &  5.7   \\
    \bottomrule
    \end{tabular}
    \caption{Parameters of the Morse component of the potential in Eq.~\eqref{eq:AgI}. For cation-cation interaction, only the Coulomb interaction is considered (i.e. $D_\mathrm{Ag,Ag} = 0$). The fractional charges $z_\mathrm{Ag} = - z_\mathrm{I} = 0.3181$ are used for the Coulomb term. }
    \label{tab:AgI}
\end{table}

\subsection{Details on MD simulations}
\label{sec:details}
All simulations are performed with the \lammps~software \cite{LAMMPS}. It has the great advantage, with respect to other MD codes, that force computation is not subject to minimum image conventions, and one can use cutoffs larger than half the simulation domain size, thanks to the inclusion of ``ghost'' atoms. This is particularly important for the purpose of this work, where small boxes are needed for the FSE analysis, yet the cutoff radius should be the same to avoid changes in the form of the potential.  
The long-range interactions are included by means of the Ewald-summation technique in MD simulations of \PbF2, \CaF2, and AgI, and with the PPPM method for the MD simulations of \UO2 \cite{Hockney2021}.
The simulations of PbF$_2$, CaF$_2$ and $\alpha$-AgI are run with a MD timestep of 4 fs. For UO$_2$, the MD time step is set to 2 fs. The trajectories ($\approx 800$ ps), from which the mean square displacements of the atoms are computed, are sampled each 10 MD time steps. For a given material, all constant-$NVT$ (canonical) and constant-$NVE$ (microcanonical) simulations ($N$ is the number of particles, $V$ the cell volume, $T$ the temperature, $E$ the total energy) are run at fixed lattice constant, $a$, irrespective of the temperature, i.e., no thermal expansion is considered for simplicity. I employ the following lattice constants: $a_\mathrm{PbF_2} = 6.056\,\mathrm{\AA}$ (value at $T=792$ K in Ref.~\onlinecite{Walker1982}); $a_\mathrm{CaF_2} = 5.712\,\mathrm{\AA}$ \cite{Dixon1980}; $a_\mathrm{UO_2} = 5.65\,\mathrm{\AA}$ (online material of Ref.~\onlinecite{Cooper2014b}); $a_\mathrm{\alpha AgI} = 5.37\,\mathrm{\AA}$ \cite{Niu2018}.
The $NVT$ simulations are run with a Bussi-Donadio-Parrinello stochastic-velocity-rescaling (SVR) thermostat \cite{Bussi2007}, as implemented in \lammps. The temperature damping parameter of the SVR thermostat is set to 100 MD timesteps.
Further details on the equilibration procedures and on the cutoffs employed are reported, for the sake of reproducibility, in the input scripts of the simulations, which are all available in the Materials Cloud repository of this work (see ``Data Availability'').

\section{Results}
\label{sec:results}

I proceed investigating FSE for physical quantities important for superionics, namely the specific isochoric heat capacity, Sec.~\ref{sec:crit_T}; the diffusivity of the mobile species, Sec.~\ref{sec:diff_coeff}; the thermal conductivity, Sec.~\ref{sec:kappa}; and the Debye-Waller $B$-factor of the non-diffusive ions of the solid matrix, Sec.~\ref{sec:DW}.

\subsection{Specific heat capacity and critical temperature}
\label{sec:crit_T}

\begin{figure*}
    \centering
    \includegraphics[width=\linewidth]{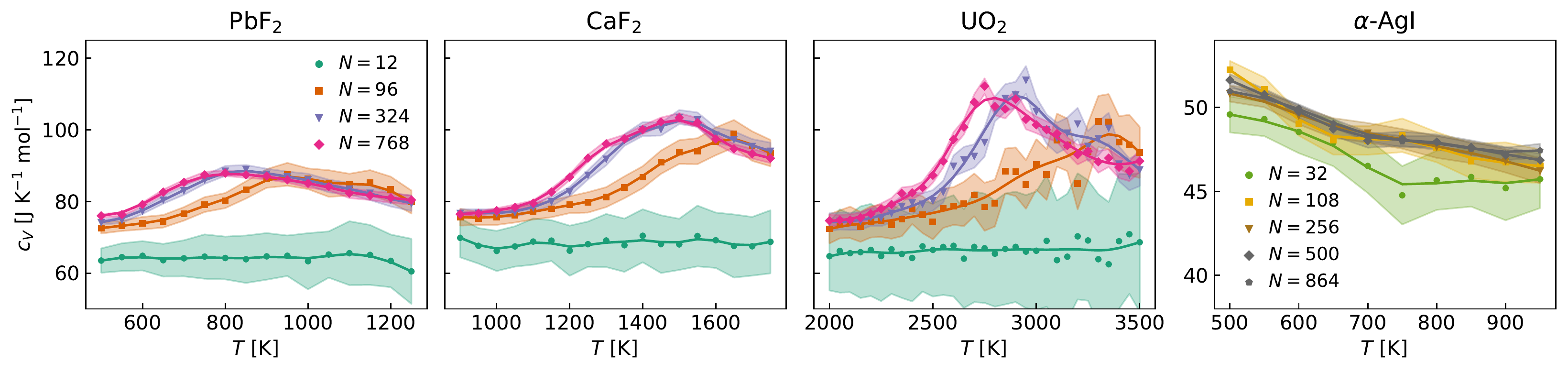}
    \caption{Isochoric specific heat capacity, $c_V$, as a function of temperature and system size for the fluorite-structure materials investigated. The markers indicate the calculated quantities, while the line is a 3 point {running-average} window filter. The shaded area indicates the statistical uncertainty, obtained from standard block analysis.}
    \label{fig:c_V}
\end{figure*}

The isochoric molar specific heat capacity, $c_V$ is obtained from the finite-difference derivative of the average energy with respect to the temperature, and displayed vs $T$ in Fig.~\ref{fig:c_V} for the four materials considered. 

Let us first focus on the fluorite-structure materials (first three panels of Fig.~\ref{fig:c_V}). 
For sufficiently large cells ($\ell \geq 2$, i.e. $N \geq 96$), the heat capacity clearly displays a peak. Experimentally, a peak in the heat capacity---at some high critical temperature, yet below the melting point---has been observed via heat-content measurements of fluorite-structure materials, and associated with a transition which is not of the 1st order; such anomaly accompanies a sensible onset of electrical conduction, indicating a transition to the superionic phase \cite{Naylor1945,Derrington1976}. At large enough size, the calculated $c_V$ is in fairly good agreement with experiments.
As mentioned in Sec.~\ref{sec:discuss}, it is known since the late '60s that the effect of a finite size is to \textit{broaden} a second-order transition and to \textit{shift} the (pseudo)critical temperature $T_c(\ell)$ with respect to its thermodynamic-limit value $T_c(\infty)$ \cite{Ferdinand1969}. 
Whether the shift is positive or negative depends, among other factors, on the boundary conditions: usually, in PBC, $T_c(\ell) > T_c(\infty)$, \textit{as a result of extra ``communication'' via paths that encircle the torus} (\textit{verbatim} from Ref.~\onlinecite{Ferdinand1969}). 
This is in fact the behaviour observed in Fig.~\ref{fig:c_V}, where the peak shifts towards lower temperatures as the size of the simulation is increased, in agreement with existing literature \cite{Yakub2007}. In the smallest cell, $\ell = 1$ and $N=12$, no peak is observed and, in the temperature range that I consider, the heat capacity is always sensibly lower than the one obtained with larger simulation boxes. 
I remark that, in agreement with Ref.~\onlinecite{Cooper2014b}, the melting of \UO2, predicted for this potential at $\approx 3100$ K \cite{Cooper2014a} via a moving interface method, is not observed in these simulations, where the lattice parameter is kept fixed to its value at $\approx 2600$ K \cite{Fink2000}, see Sec.~\ref{sec:details}.

In striking contrast with fluorite structure materials, for $\alpha$-AgI no significant size effect is observed, in line with its intrinsic superionic structure.
The link between FSE and the onset of a superionic phase transition is clearly highlighted from the analysis of the diffusivity of the mobile species, as discussed below.

\subsection{Diffusion coefficient}
\label{sec:diff_coeff}

The diffusion coefficient of the mobile species is the most characterizing quantity of superionic conductors: its behavior at the superionic transition dictates the classification of SI materials (see Fig.~\ref{fig:fluorites_AgI}). 
Ion diffusion is usually described as an Arrhenius-like process
\begin{equation}
    D(T) = A \exp\left[ - \frac{E_a}{kT} \right]
    \label{eq:arrhenius}
\end{equation}
where {$k$ is Boltzmann's constant,} $A$ is a prefactor with the dimensions of a diffusivity, and $E_a$ is the activation energy of the hopping mechanism leading to particle diffusion. In SI materials, in general, for a specific regime of diffusion, $A$ and $E_a$ weakly depend on the temperature: it makes sense, therefore, to plot the logarithm of the diffusion coefficient against the inverse temperature (in the so-called Arrhenius plot) to highlight significant changes in the diffusion mechanism. $E_a$ represents the slope of the Arrhenius plot, and $A$ its intercept.
\begin{figure*}
    \centering
    \includegraphics[width=\linewidth]{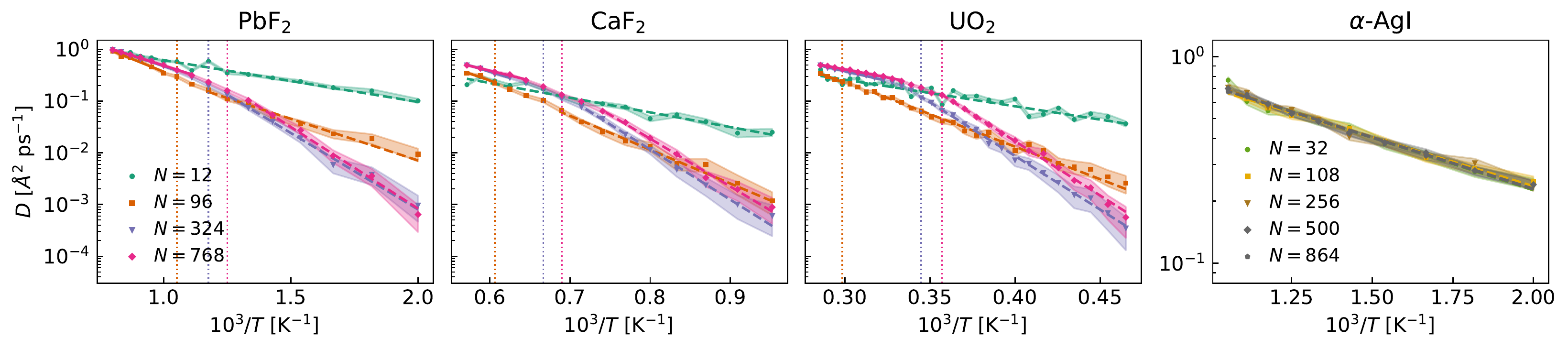}
    \caption{Arrhenius plot of the diffusion coefficient of the mobile species for the fluorite-structure materials considered in this works, as a function of the cell size. The markers indicate the calculated $D$ and the shaded area the uncertainty from a block analysis on 4 blocks of $\approx 170$ ps each. In the plots for fluorite-structure materials, the solid (dashed) lines are fits to Eq.~\eqref{eq:arrhenius} above (below) $T_c(\ell)$. The dotted vertical lines indicate the size-dependent critical temperature, $T_c(\ell)$, obtained from the position of the maximum of the heat capacity (same color code).}
    \label{fig:D}
\end{figure*}

Figure \ref{fig:D} displays the Arrhenius plots of the diffusivity of the mobile species for the materials considered in this work, at different system sizes.
The reported values for $D$ (markers) are obtained from the slope of the mean square displacement (MSD) of the mobile species at large enough time:
\begin{equation}
    D = \frac{1}{6} \lim_{t\to\infty}\frac{d}{dt} \frac{1}{N_d} \sum_{i=1}^{N_d} \left\langle \left|\mathbf{r}_i(t) - \mathbf{r}_i(0)\right|^2 \right\rangle,
\end{equation}
$N_d$ being the number of atoms of the diffusive species, and $\mathbf{r}_i(t)$ the position of the $i$-th atom of the mobile species at time $t$. A linear fit on the MSD at large $t$ was used to obtain $D$.
A block analysis, with 4 blocks of $\approx 170$ ps each, is performed to extract the uncertainty on the MSD{. This uncertainty propagates to the} asymptotic-time slope {of the MSD}, and eventually to the diffusion coefficient. The uncertainty on $D$ is reported as the shaded area in the Arrhenius plots. The reported $D$ is computed in the ``laboratory'' reference frame, where the center of mass of the non-mobile species is fixed, although the MD simulations are performed in the barycentric reference frame, where the \textit{total} momentum vanishes. The relations between the different, reference-frame dependent definitions of the diffusion coefficients are described in the Supplementary Material \cite{SupplMat}, Sec.~S2.A. To check whether the SVR thermostat affects the estimate of $D$, I also run $NVE$ simulations, previously equilibrated at the target temperature. Diffusion coefficients extracted from $NVE$ and $NVT$ simulations display fully compatible Arrhenius plots for each simulation cell size, as reported in the Supplementary Material \cite{SupplMat}, Fig.~S9.

In \PbF2, \CaF2, and \UO2, I find that $D(T)$ strongly depends on the system size. For instance, the diffusivity for $N=96$ is greater (smaller) than the large-$N$ values at low (high) $T$, in agreement with previous observations \cite{Dixon1980}. 
Furthermore, the kink in the Arrhenius plot at $T_c$, characterizing type II materials, is observable only for $N\geq 324$ (i.e. $\ell\geq 3$). This is quantified by the values of $E_a(\ell)$ (the slope of the Arrhenius plot), extracted from the fit of $D$ to Eq.~\eqref{eq:arrhenius} below and above the size-dependent critical temperature, $T_c(\ell)$, and reported in Table \ref{tab:E_a}: only for $N\geq 324$ a sensible difference between $E_a (T<T_c)$ and $E_a (T>T_c)$ is observed. 
Notice, as well, that the kink in the Arrhenius plot, associated to the SI transition, moves towards lower temperatures as the size is increased, in accordance to the existing literature \cite{Potashnikov2013}, and with the shift in $T_c$ extracted from the maximum of the heat capacity, Fig.~\ref{fig:c_V}. 

As shown in Supplementary Material \cite{SupplMat} Figs.~S4 and S5, all these results are in qualitative agreement also with frozen-matrix simulations \cite{Kahle2018}, where the non-diffusive ions of the solid matrix are kept fixed to their equilibrium position, and with simulations employing a short-range version of the Coulomb interaction \cite{Fennell2006}. This agreement confirms the proposed picture whereby the FSE on the diffusivity of mobile ions are mainly imputable to geometric factors (like the degeneracy of empty sites, and the extra interatomic communication paths occurring in PBC), rather than to the details of the potential or the vibrations of the solid matrix. 
{This is also justified by a set of simulations run on non-stoichiometric lead fluoride, where a size-independent concentration of empty sites is generated by removing a set of randomly selected F$^-$ ions accordingly: the presence of available empty sites favors hopping on a more local scale than in systems with perfect stoichiometry, where, below the SI transition, all the regular sites are occupied and the hopping of one F$^-$ anion can only occur with a net hopping of other F$^-$ anions, and results in a drastic reduction of FSE on fluorine-ion diffusion, as shown quantitatively in Appendix \ref{app:empty_sites}.}

\begin{table}[]
    \centering
    \begin{tabular}{
    c
    c
    S[table-format=4.0(4)]
    S[table-format=4.0(4)]
    S[table-format=4.0(4)]}
      $N=$   & 
      {$12$} & 
      {$96 $}& 
      {$324$} & 
      {$768$} \\
      \toprule
      \multirow{2}*{\PbF2} & \multirow{2}*{${160 \pm 9}$} &
      320 \pm 20 &
      590 \pm 20 &
      620 \pm 20 \\
      & 
      &
      390 \pm 40 &
      324 \pm 11 &
      308 \pm 8 \\
      \midrule
      \multirow{2}*{\CaF2}   & \multirow{2}*{${560 \pm 50}$}
      &
      1300 \pm 60 &
      1930 \pm 40 &
      1830 \pm 60 \\
      &
      &
      1070 \pm 330 &
      940 \pm 70 &
      750 \pm 40 \\
      \midrule
      \multirow{2}*{\UO2}   & \multirow{2}*{${1020 \pm 80}\;$} &
      2490 \pm 70 &
      4110 \pm 80 & 
      4160 \pm 50 \\
      &
      &
      2410 \pm 150 &
      1490 \pm 70 &
      1240 \pm 40 \\
      \bottomrule
    \end{tabular}
    \caption{Activation energies, $E_a$, (in meV) of the considered fluorite-structure materials, as a function of the system size, from weighted fit to Eq.~\eqref{eq:arrhenius}. For each system, the first row corresponds to the regime $T<T_c(\ell)$, and the second row to the regime $T>T_c(\ell)$. For $N=12$ (i.e., $\ell=1$), where no phase transition is observed, only one value is provided. See the Materials Cloud Repository for data and details of the fit.}
    \label{tab:E_a}
\end{table}

In striking contrast with fluorite-structure materials, in $\alpha$-AgI, the values of the diffusivity of Ag ions at different cell sizes (even for very small cells) are all consistent with each other. The activation energies range from $93 \pm 3$ to $100 \pm 7$ meV (without any particular trend connected to the box size), and agree with the experimental value $E_a = 94.97$ meV of Ref.~\onlinecite{Kvist1970}. 
As suggested above, I ascribe the size independence of the diffusivity to the large degeneracy of equivalent Ag ion sites within the conventional unit cell of $\alpha$-AgI: in this superionic phase, Ag-ion hopping occurs on a smaller length scale than in fluorite-structure materials, where, instead, the hopping of one ion can occur only if accompanied by the concerted hopping of other ions, since---apart from the short transient of the jump---all the regular sites are occupied, as confirmed also in the literature (see e.g. Ref.~\onlinecite{Walker1982, Gillan1980}). 

\subsection{Thermal conductivity}
\label{sec:kappa}

\begin{figure*}
    \centering
    \includegraphics[width=\linewidth]{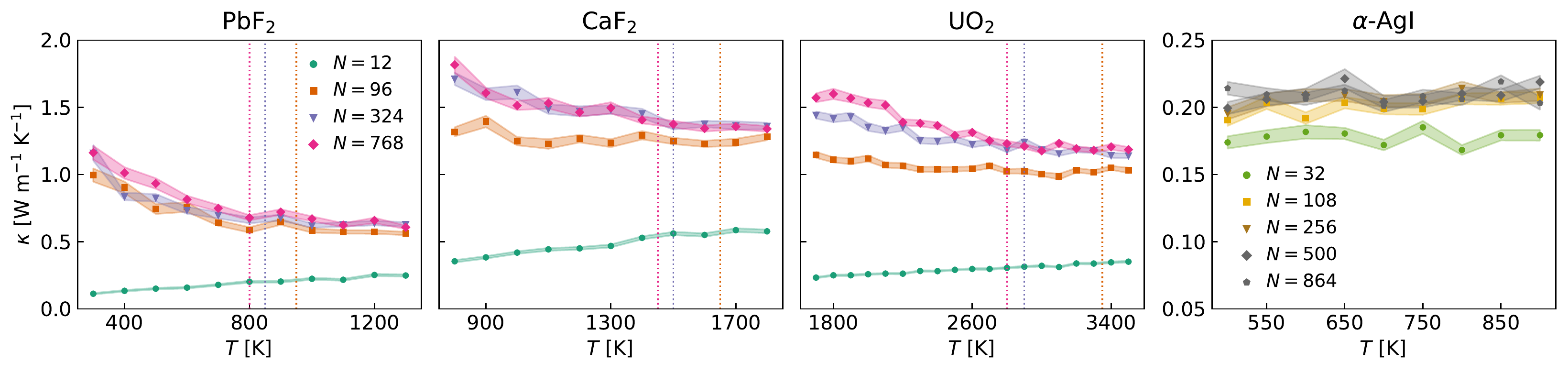}
    \caption{Temperature behavior of thermal conductivity (markers) and associated uncertainty (shaded area) from cepstral analysis. The vertical dotted lines indicate the size-dependent critical temperature to the SI phase, obtained from the position of the maximum of the specific heat capacity, see Fig.~\ref{fig:c_V}.}
    \label{fig:kappa}
\end{figure*}

Figure \ref{fig:kappa} shows the temperature and size dependence of the thermal conductivity, $\kappa$, for the materials considered, extracted from $NVT$ MD simulations according to the Green-Kubo theory of linear response for multicomponent, diffusive systems, as described in Appendix \ref{app:details_kappa}. Even if a mode-based analysis \cite{Simoncelli2019,Isaeva2019} of heat conduction would be required for a quantitative assessment of the role of disorder on phonon propagation, here I employ the GK formula and MD simulations to include the role of diffusing ions, which, by construction, is not considered in mode-based calculations.

Let us first focus, again, on fluorite-structure materials. At large $T$, $\kappa$ is almost independent of $T$ due to the increasing disorder that accompanies the diffusion of mobile ions, suppressing phonon propagation. This is a manifestation of the Allen-Feldman regime \cite{AllenFeldman}, and a typical feature observed also in other SI materials, like solid-state electrolytes \cite{Pegolo2021}.
The FSE of $\kappa$ are system dependent, the values for \PbF2 being almost converged for $N=96$, in contrast with \CaF2 and \UO2. Nonetheless, a general trend can be observed.
The $N=12$ $(\ell=1)$ cell dramatically fails in describing heat transport, and displays a non-physical, slight increase of $\kappa$ with temperature.
For $\ell>1$, the Allen-Feldman limit is achieved at \textit{lower temperatures for smaller simulation boxes}, in agreement with the larger diffusivity at low $T$ observed for small size, and the higher degree of disorder in smaller cells \footnote{This is also confirmed by the temperature, lower for smaller systems, at which the multicomponent analysis departs from the single-component one, which assumes no atomic diffusion (see Fig.~\ref{fig:kappa_multi_single}).}. 
Furthermore, the presence of a single defect is not supposed to strongly perturb a large system, while it would dramatically affect a few-atom cell, suppressing phonon propagation in favor of the Allen-Feldman limit.
The activation of diffusion with the onset of significant disorder is much more relevant for heat transport than the actual SI phase transition: at $T=T_c(\ell)$ (vertical dotted lines) no particular feature of $\kappa(T)$ is in fact observed.

Overall, the large-cell values are in good agreement with existing literature: $\kappa_\mathrm{PbF_2} = 1.4\,\mathrm{W m^{-1} K^{-1}}$ at 300 K, Ref.~\onlinecite{Popov2017} (experimental); $\kappa_\mathrm{CaF_2} = 1.46 \pm 0.29 \, \mathrm{W m^{-1} K^{-1}}$ at $T=1694$ K, Ref.~\onlinecite{Lindan1991} (numerical simulation, same density and potential); $\kappa_\mathrm{UO_2} = 1.5 \, \mathrm{W m^{-1} K^{-1}}$ at $T=3000$ K, Ref.~\onlinecite{Lindan1991} (numerical simulation, close density but different potential). I remark that, in \UO2, for $T \gtrsim 2000$ K, the growing contribution of electrons to heat conduction must be added for a comparison with experiments. 

The thermal conductivity of $\alpha$-AgI is almost constant in the considered temperature range. This indicates that phonon propagation is always suppressed by disorder in favor of the AF regime in this intrinsically superionic phase. All simulations with $N \geq 108$ are fully compatible. 
For the small cell, $N=32$, $\kappa$ is lower, though only by 10-15\%, than the converged value. The results are in good agreement with the experimental value of $0.17$ $\mathrm{W m^{-1} K^{-1}}$ at $T\approx 500$ K \cite{Goetz1982}.

\subsection{Dynamics of the non-diffusive species}
\label{sec:DW}

FSE affect also the dynamics and thermal vibrations of the solid matrix, which I investigate in this section in terms of the mean-square-displacement of the non-diffusive (n.d.)~ions with respect to their equilibrium position $  \langle \mathbf{u}^2 \rangle_{\mathrm{n.d.}} \equiv \langle[\mathbf{r}(t) - \mathbf{r}_{\mathrm{eq}}]^2\rangle_{\mathrm{n.d.}}  = \tfrac{1}{2} \lim_{t\to\infty} \langle \left|\mathbf{r}(t) - \mathbf{r}(0)\right|^2 \rangle_\mathrm{n.d.}$, which is a time-independent quantity. Furthermore, following a standard convention, I shall employ the so-called $B$-factor, $B \equiv \tfrac{8\pi^2}{3} \left\langle \mathbf{u}^2 \right\rangle_\mathrm{n.d.}$, entering the Debye-Waller factor that dictates the attenuation of X-ray or neutron scattering in experiments.

\begin{figure}
    \centering
    \includegraphics[width=0.95\linewidth]{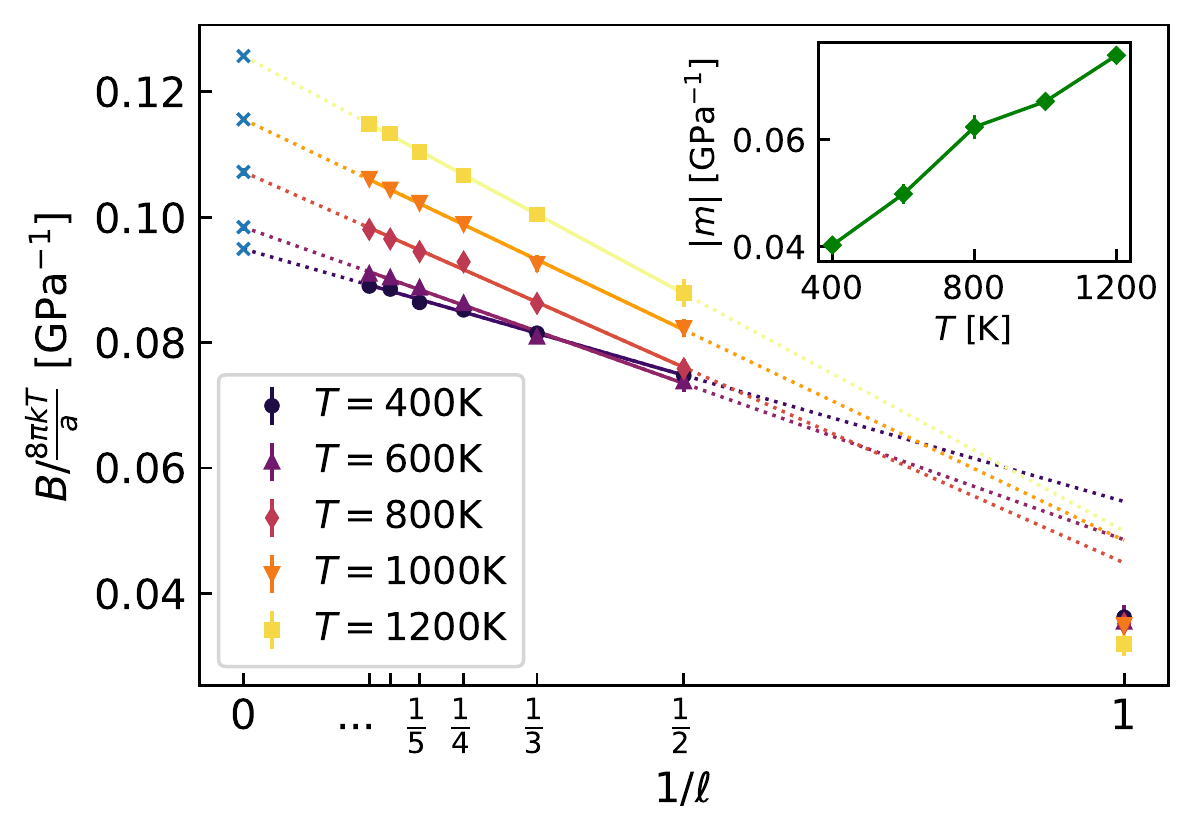}
    \caption{Finite-size effects affecting the Debye-Waller $B$-factor of Pb atoms in \PbF2, at different temperatures. Values obtained from simulations (markers and error bars), and the fit $B(\ell) = m/\ell + B(\infty)$ (for $\ell\geq 2$) are displayed. The blue crosses indicate the extrapolated asymptotic value $B(\infty)$. The absolute value of the slope, $|m|$, is reported in the inset. The linear behavior in $1/\ell$ is evident. For $\ell=1$ higher order FSE may occur. }
    \label{fig:B_PbF2_DW}
\end{figure}

For SI phases, MD simulations are needed to compute the $B$-factor, since normal-mode-based approaches \cite{Malica2019} cannot be applied. Nonetheless, the values obtained from MD simulations are known to slowly converge with size. Early calculations for cubic hard-sphere solids show FSE corrections that follow a $1/\ell$ law \cite{Young1974}. This law stems from the minimum frequency which can be sampled in a finite box size. An extensive derivation of this FSE is provided in the Supplementary Material, with an example on FCC solid argon. 
Interestingly, the same trend is observed in my simulations on SI materials. 
Figure~\ref{fig:B_PbF2_DW} shows the $B$-factor of \PbF2, as a function of $1/\ell$, at different temperatures. The linear behavior is evident, and a linear fit of the data can be used to extrapolate the $B$-factor for infinite size, $B(\infty)$ (blue crosses). Notice that the $B$-factor is here re-scaled by $T$ and other geometric factors so that the slope represents the inverse of an effective elastic modulus of the material (see Supplementary Material, Sec.~5.A). Only the smallest box considered (i.e., the conventional unit cell, $\ell=1$, with $N=12$ atoms) deviates from the $1/\ell$ behavior, not surprisingly, as it fails in describing most of the quantities analyzed so far, and where higher-order FSE may enter. 

\begin{figure*}
    \centering
    \includegraphics[width=\linewidth]{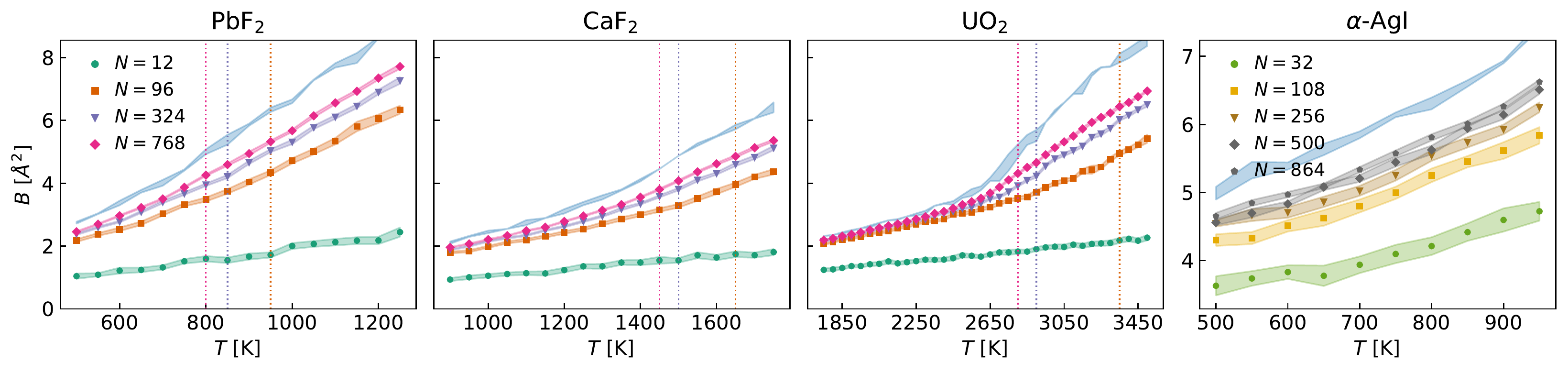}
    \caption{$B$-factors of the non-diffusive ions for the materials considered in this work, as a function of the system size. The vertical dotted lines indicate the size-dependent critical temperature to the SI phase, obtained from the position of the maximum of the specific heat capacity. The blue, shaded area represents the extrapolated value, accounting for the finite-size correction.
    }
    \label{fig:Bfactors}
\end{figure*}

The $1/\ell$ behavior is observed in all the systems considered. An exception is the temperature range between 2600 and 2900 K for \UO2: in fact, the drastic change in the elastic properties of \UO2, and therefore of $B$, at the SI phase transition, occurs at different critical temperature for different sizes (see Sec.~\ref{sec:crit_T}).
Figure~\ref{fig:Bfactors} displays the $B$-factor as a function of temperature for different systems and sizes. The blue, shaded area represents $B(\infty)$ with its uncertainty. Notice that significant FSE are here observed not only in fluorite-structure materials, but also in $\alpha$-AgI. In fact, the very reason why the $B$-factor exhibits FSE---the minimum mode frequency which can be probed for a given cell---is very general and not system dependent. 
Once again, these calculations show that wrong results may be obtained even from an accurate description of interatomic interactions in a MD simulation (e.g. by computing forces \textit{ab initio}), whenever FSE are not correctly accounted for.

\section{Conclusions}
\label{sec:conclusions}
Finite size effects (FSE) in superionic materials are in general not negligible, and strongly depend on the specific system and property examined.
In materials that possess intrinsically a high availability of degenerate hopping sites, like $\alpha$-AgI, the FSE on the diffusivity---the key quantity of superionic conductors---are weak. 
Things change, instead, whenever the mechanism of diffusion strongly depends on the temperature range, like in fluorites. These systems display an order-disorder transition to the superionic regime without a net change neither in the lattice structure nor in the available hopping sites, which coincide with those occupied by the mobile ions in the non-superionic phase.
This work shows that, in these materials, the diffusivity is strongly affected by the system size, even qualitatively: the change of activation energy of the diffusion process at the SI critical temperature, resulting in the typical kink in the Arrhenius plot of the diffusivity, can be observed only above a certain threshold size. The order-disorder critical temperature is also largely affected by FSE, being larger for smaller sizes, as indicated by the shift of the maximum of the specific heat capacity. Therefore, in these materials, the tangling between hydrodynamics (diffusion processes) and thermodynamics (the specific phase of a material) is responsible for changes in the diffusivity of even some orders of magnitude, at a given temperature, depending on the size.
The thermal conductivity is also affected by FSE, mainly due to the role of disorder in hindering phonon propagation: this effect is larger at smaller simulation boxes, where the Allen-Feldman regime is reached at lower temperatures. The transition to the SI phase does not seem to strongly affect $\kappa$, instead.
{In general, an \textit{a priori} determination on the minimum size that is sufficient to obtain satisfactory results is hardly feasible, due to strong dependence of FSE on the specific superionic material. My analysis suggests, \textit{a posteriori}, that, for fluorite-structure materials, a minimum of $N=324$ atoms is needed to correctly capture at least the main qualitative features of particle diffusion and heat transport, while for the $\alpha$-AgI phase, even a relatively small cell of $N=108$ atoms is sufficient to obtain a quantitative convergence of all the analyzed properties.}

FSE also affect the Debye-Waller $B$-factor incorporating the thermal motion of the non-diffusive ions of the solid matrix. The $1/\ell$ (or $1/N^{1/3}$) behavior, predicted for simple solids, is observed also for the SI materials considered. This is motivated by the minimum frequency of oscillation which can be captured by a simulation of a given size, irrespective of the specific system or phase considered.

Increasingly reliable interatomic potentials have provided, during the last few years, a systematically more accurate description of SI materials. However, this work clearly shows that accurate potentials alone are not sufficient to converge the calculation of many key properties, if not complemented with a full treatment of FSE. From the application point of view, particular attention must be paid in calculations aiming to compare different candidates for realistic devices, like solid-state batteries. A given simulation size may be sufficient for some superionic materials but not for others.

\section*{Supplementary Material}
See supplementary material\cite{SupplMat} for more details about \textit{i}) the calculation of the heat capacity; \textit{ii}) the calculation of the diffusivity (role of the reference frame, of modified Coulomb interactions and of vibrations of the solid matrix); \textit{iii}) a comparison of the results obtained with \textit{NVT} and \textit{NVE} simulations; \textit{iv}) the calculation of the thermal conductivity; \textit{v}) the derivation of the $1/\ell$ law for FSE of the $B$-factor and a simple application to solid argon; \textit{vi}) the heat capacity of the defected structure of Appendix \ref{app:empty_sites}.

\section*{Data Availability}
The data that support the plots and relevant results within this paper are available on the Materials Cloud platform\cite{talirz2020materials}. See DOI: \href{https://doi.org/10.24435/materialscloud:jy-tw}{https://doi.org/10.24435/materialscloud:jy-tw}.

\section*{Author Declarations}
The author has no conflicts to disclose.

\section*{Acknowledgements}

I thank Michele Ceriotti, Loris Ercole, Alfredo Fiorentino, Lorenzo Gigli and Paolo Pegolo for insightful discussions and fruitful comments on the manuscript. I acknowledge funding from the European Union's Horizon 2020 research and innovation programme under the Marie Sk\l{}odowska-Curie Action IF-EF-ST, grant agreement no. 101018557 (TRANQUIL).

\appendix

\section{Thermal conductivity for diffusive, multicomponent systems}\label{app:details_kappa}
The thermal conductivity, $\kappa$, is the proportionality coefficient between the energy flux and the (negative of the) temperature gradient \textit{in the absence of any convection}.
For a two-component system, like the SI materials studied in this work, characterized by the presence of one diffusive species \cite{Galamba2007, Pegolo2021}, the Green-Kubo theory of linear response allows to extract $\kappa$ from MD simulations as the zero frequency component of:
\begin{equation}
    \kappa(\omega) = \frac{V}{6kT^2} \left[ S_{ee}(\omega) - S_{de}(\omega) S_{dd}^{-1}(\omega) S_{ed}(\omega) \right]  \label{eq:kappa}
\end{equation}
where 
\begin{equation}
    S_{AB}(\omega) = \int_{-\infty}^{+\infty} e^{i\omega t} \langle \mathbf{J}_A (t) \cdot \mathbf{J}_B(0) \rangle dt
\end{equation}
is the power spectrum of the fluxes $\mathbf{J}_A(t)$ and $\mathbf{J}_B(t)$. The flux $\mathbf{J}_e(t)$ is the total energy flux, here computed via the \texttt{compute heat/flux} command of \lammps, while $\mathbf{J}_d(t) = \frac{1}{V} \sum_{i=1}^{N_d} \dot{\mathbf{r}}_i $ is the convective flux of the diffusive species. The choice of the laboratory or the barycentric reference frames is irrelevant, provided that both $\mathbf{J}_e$ and $\mathbf{J}_d$ are computed in the same reference frame. A multivariate technique \cite{Bertossa2019,baroni2018} for the analysis of time-series of the energy flux obtained from MD simulations has been employed in this work, allowing one to compute $\kappa$, efficiently and rigorously, for multicomponent superionic materials. As for the diffusivity, I checked any possible influence of the thermostat on the value of $\kappa$, by repeating several simulations in the $NVE$ ensemble, previously equilibrated at the desired temperature. The results from $NVT$ and $NVE$ are fully compatible (see Supplementary Material, Fig.~S10 \cite{SupplMat}).

\begin{figure}
    \centering
    \includegraphics[width=0.9\linewidth]{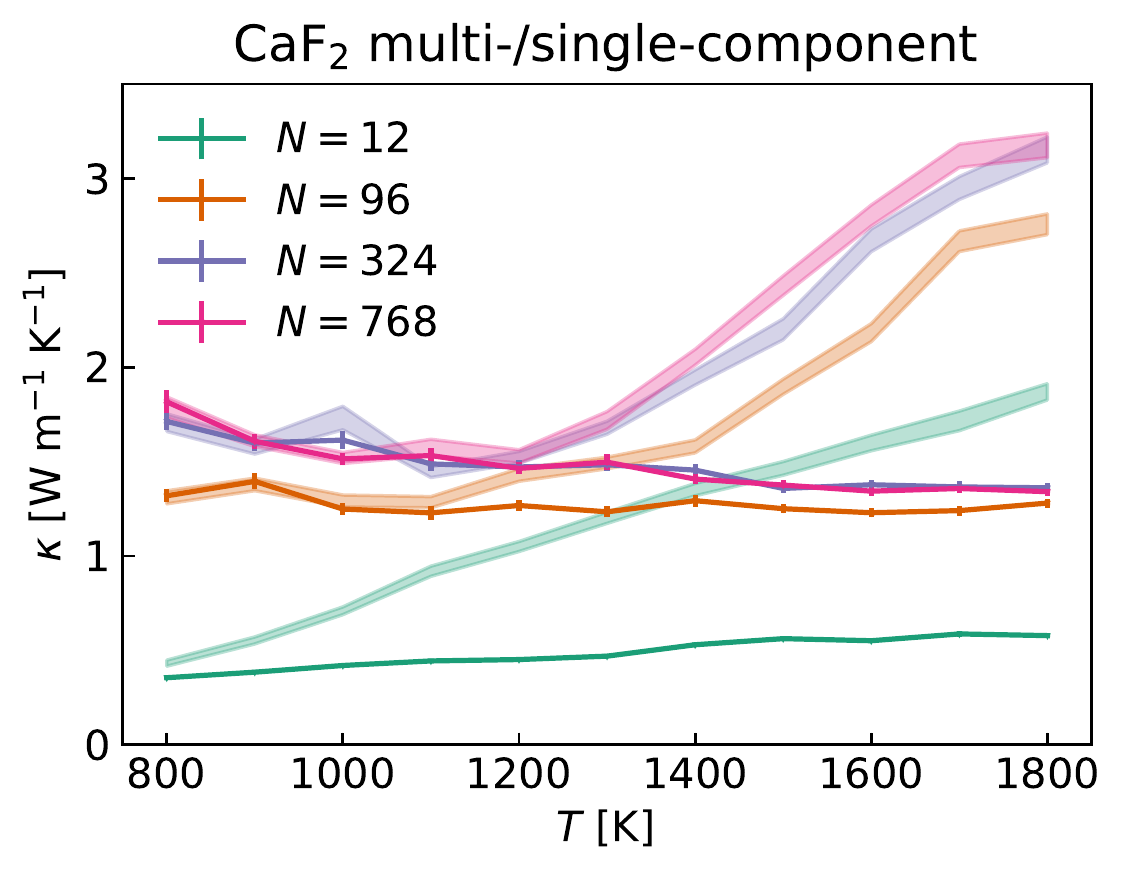}
    \caption{Comparison between multicomponent (lines and errorbars) and single-component (shaded areas) calculation of the thermal conductivity, $\kappa$, as a function of temperature and system size, for \CaF2.}
    \label{fig:kappa_multi_single}
\end{figure}

The second term between square brackets in Eq.~\eqref{eq:kappa} represents the contribution to heat flux due to \textit{convection}. It must be removed from the first term, $S_{ee}$ to correctly calculate $\kappa$, i.e.~the coefficient of thermal \textit{conduction} \cite{Lindan1991,Galamba2007,Bonella2017,Grasselli2021}. Its effects on the thermal conductivity of \CaF2 at different temperatures and system sizes are shown in Fig.~\ref{fig:kappa_multi_single}: the full calculation, employing Eq.~\eqref{eq:kappa}, (solid lines and errorbars) coincides with the single component calculation (shaded areas),  that is $\kappa = \frac{V}{6kT^2} S_{ee}$, only at low temperatures, where diffusion is negligible. Notice that the departure from the multicomponent value occurs at lower temperature for small sizes, in agreement with overestimation of $D$ in small boxes at low $T$.

\section{Role of additional F$^{-}$ empty sites on diffusion}
\label{app:empty_sites}

\begin{figure}[h!]
    \centering
    \includegraphics[width=\columnwidth]{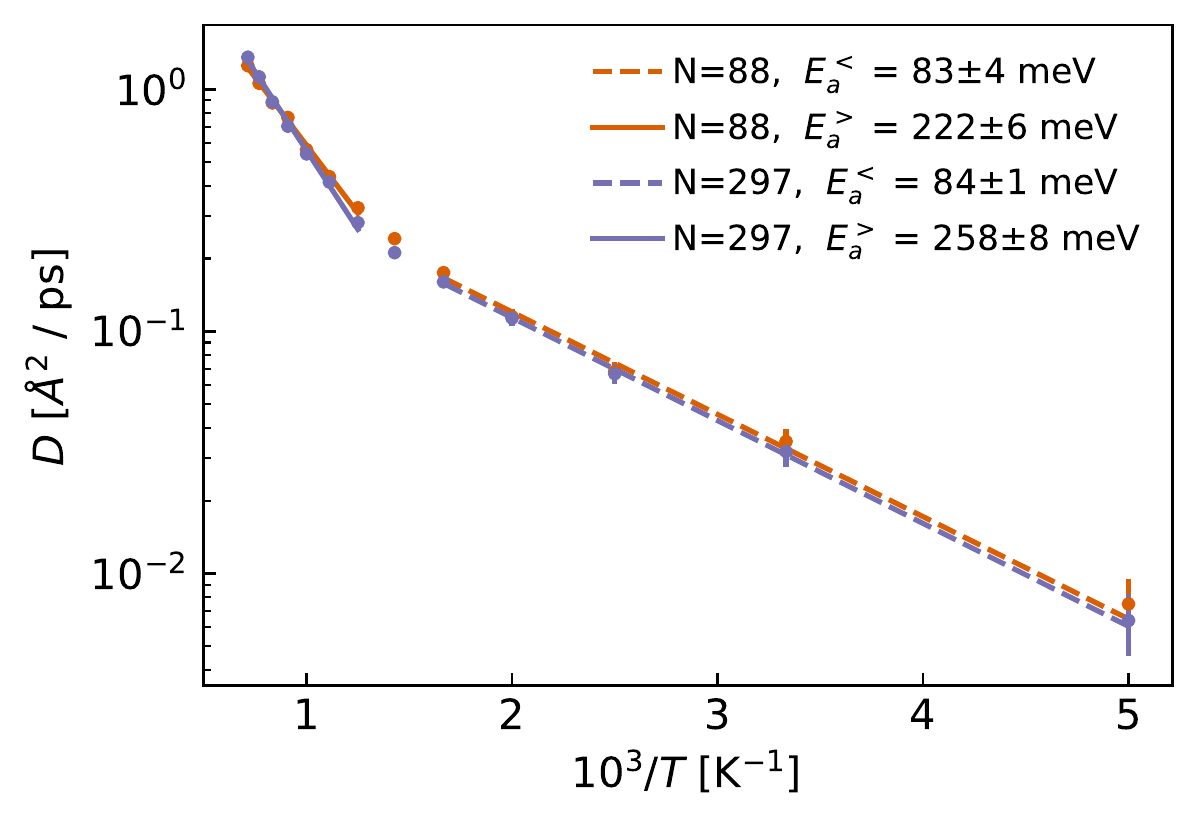}
    \caption{{Arrhenius plot of the diffusivity of F$^-$ ions at different sizes, for systems where a number $\ell^3$ of F$^-$ ions are randomly selected and removed from the stoichiometric system. The activation energies for the diffusion process below ($E_a^<$) and above ($E_a^>$) the transition temperature to the SI regime are extracted from the fits to Eq.~\eqref{eq:arrhenius} and reported in the legend.} }
    \label{fig:PbF2_defected}
\end{figure}

{I investigated the role of a higher number of accessible empty sites, to which a F$^-$ ion can jump, by considering a strongly defected lead fluoride system where, for each simulation, $\ell^3$ fluorine ions are randomly selected and removed from the $\ell\times\ell\times\ell$ supercell with perfect stoichiometry, $\ell$ being, as usual, the number of replicas of the conventional cell. In this way, the concentration of F$^-$ vacancies is set to 1/8, and it is independent of $\ell$. Therefore, the simulation cells possess $4\ell^3$ lead ions and $7\ell^3$ fluorine ions. The selection of the F$^-$ ions to remove, and the generation of resulting defected cell, is performed with the code \textsc{atomsk} \cite{atomsk}. A different seed for the random selection is used for each temperature and size. After a long initial equilibration (400 ps), a production run in the $NVT$ ensemble is performed with \lammps. A uniform, neutralizing, background charge distribution is implicitly applied in the simulations employing Ewald's method.} 

{Although such a high concentration of F$^-$ vacancies is hardly attainable in actual experimental samples, this model serves as a test bench to validate the general picture whereby the presence of empty sites tends to reduce FSE, as extensively discussed in the main manuscript. In fact, as shown in Fig.~\ref{fig:PbF2_defected}, in these simulations, convergence in the Arrhenius plots is reached already for the $\ell =2$ cell ($N=88$ atoms), in striking contrast to cells with perfect stoichiometry. Another interesting difference with the perfect stoichiometry case is that the activation energy below the transition to the superionic phase is significantly \textit{lower} than that above the SI transition (whose value is itself quite close to the one for perfect stoichiometry): the presence of empty sites facilitates F$^-$ ion diffusion below the transition to the SI phase, while it becomes much less relevant above it, where the system is globally disordered.} 

\FloatBarrier


\begin{thebibliography}{82}%
\makeatletter
\providecommand \@ifxundefined [1]{%
 \@ifx{#1\undefined}
}%
\providecommand \@ifnum [1]{%
 \ifnum #1\expandafter \@firstoftwo
 \else \expandafter \@secondoftwo
 \fi
}%
\providecommand \@ifx [1]{%
 \ifx #1\expandafter \@firstoftwo
 \else \expandafter \@secondoftwo
 \fi
}%
\providecommand \natexlab [1]{#1}%
\providecommand \enquote  [1]{``#1''}%
\providecommand \bibnamefont  [1]{#1}%
\providecommand \bibfnamefont [1]{#1}%
\providecommand \citenamefont [1]{#1}%
\providecommand \href@noop [0]{\@secondoftwo}%
\providecommand \href [0]{\begingroup \@sanitize@url \@href}%
\providecommand \@href[1]{\@@startlink{#1}\@@href}%
\providecommand \@@href[1]{\endgroup#1\@@endlink}%
\providecommand \@sanitize@url [0]{\catcode `\\12\catcode `\$12\catcode
  `\&12\catcode `\#12\catcode `\^12\catcode `\_12\catcode `\%12\relax}%
\providecommand \@@startlink[1]{}%
\providecommand \@@endlink[0]{}%
\providecommand \url  [0]{\begingroup\@sanitize@url \@url }%
\providecommand \@url [1]{\endgroup\@href {#1}{\urlprefix }}%
\providecommand \urlprefix  [0]{URL }%
\providecommand \Eprint [0]{\href }%
\providecommand \doibase [0]{http://dx.doi.org/}%
\providecommand \selectlanguage [0]{\@gobble}%
\providecommand \bibinfo  [0]{\@secondoftwo}%
\providecommand \bibfield  [0]{\@secondoftwo}%
\providecommand \translation [1]{[#1]}%
\providecommand \BibitemOpen [0]{}%
\providecommand \bibitemStop [0]{}%
\providecommand \bibitemNoStop [0]{.\EOS\space}%
\providecommand \EOS [0]{\spacefactor3000\relax}%
\providecommand \BibitemShut  [1]{\csname bibitem#1\endcsname}%
\let\auto@bib@innerbib\@empty
\bibitem [{\citenamefont {Goodenough}\ and\ \citenamefont
  {Kim}(2010)}]{Goodenough2010}%
  \BibitemOpen
  \bibfield  {author} {\bibinfo {author} {\bibfnamefont {J.~B.}\ \bibnamefont
  {Goodenough}}\ and\ \bibinfo {author} {\bibfnamefont {Y.}~\bibnamefont
  {Kim}},\ }\bibfield  {title} {\enquote {\bibinfo {title} {Challenges for
  rechargeable {Li} batteries},}\ }\href@noop {} {\bibfield  {journal} {\bibinfo
  {journal} {Chemistry of materials}\ }\textbf {\bibinfo {volume} {22}},\
  \bibinfo {pages} {587--603} (\bibinfo {year} {2010})}\BibitemShut {NoStop}%
\bibitem [{\citenamefont {Janek}\ and\ \citenamefont
  {Zeier}(2016)}]{Janek2016}%
  \BibitemOpen
  \bibfield  {author} {\bibinfo {author} {\bibfnamefont {J.}~\bibnamefont
  {Janek}}\ and\ \bibinfo {author} {\bibfnamefont {W.~G.}\ \bibnamefont
  {Zeier}},\ }\bibfield  {title} {\enquote {\bibinfo {title} {A solid future
  for battery development},}\ }\href@noop {} {\bibfield  {journal} {\bibinfo
  {journal} {Nature Energy}\ }\textbf {\bibinfo {volume} {1}},\ \bibinfo
  {pages} {1--4} (\bibinfo {year} {2016})}\BibitemShut {NoStop}%
\bibitem [{\citenamefont {Kato}\ \emph {et~al.}(2016)\citenamefont {Kato},
  \citenamefont {Hori}, \citenamefont {Saito}, \citenamefont {Suzuki},
  \citenamefont {Hirayama}, \citenamefont {Mitsui}, \citenamefont {Yonemura},
  \citenamefont {Iba},\ and\ \citenamefont {Kanno}}]{Kato2016}%
  \BibitemOpen
  \bibfield  {author} {\bibinfo {author} {\bibfnamefont {Y.}~\bibnamefont
  {Kato}}, \bibinfo {author} {\bibfnamefont {S.}~\bibnamefont {Hori}}, \bibinfo
  {author} {\bibfnamefont {T.}~\bibnamefont {Saito}}, \bibinfo {author}
  {\bibfnamefont {K.}~\bibnamefont {Suzuki}}, \bibinfo {author} {\bibfnamefont
  {M.}~\bibnamefont {Hirayama}}, \bibinfo {author} {\bibfnamefont
  {A.}~\bibnamefont {Mitsui}}, \bibinfo {author} {\bibfnamefont
  {M.}~\bibnamefont {Yonemura}}, \bibinfo {author} {\bibfnamefont
  {H.}~\bibnamefont {Iba}}, \ and\ \bibinfo {author} {\bibfnamefont
  {R.}~\bibnamefont {Kanno}},\ }\bibfield  {title} {\enquote {\bibinfo {title}
  {High-power all-solid-state batteries using sulfide superionic conductors},}\
  }\href@noop {} {\bibfield  {journal} {\bibinfo  {journal} {Nature Energy}\
  }\textbf {\bibinfo {volume} {1}},\ \bibinfo {pages} {1--7} (\bibinfo {year}
  {2016})}\BibitemShut {NoStop}%
\bibitem [{\citenamefont {Kwade}\ \emph {et~al.}(2018)\citenamefont {Kwade},
  \citenamefont {Haselrieder}, \citenamefont {Leithoff}, \citenamefont
  {Modlinger}, \citenamefont {Dietrich},\ and\ \citenamefont
  {Droeder}}]{Kwade2018}%
  \BibitemOpen
  \bibfield  {author} {\bibinfo {author} {\bibfnamefont {A.}~\bibnamefont
  {Kwade}}, \bibinfo {author} {\bibfnamefont {W.}~\bibnamefont {Haselrieder}},
  \bibinfo {author} {\bibfnamefont {R.}~\bibnamefont {Leithoff}}, \bibinfo
  {author} {\bibfnamefont {A.}~\bibnamefont {Modlinger}}, \bibinfo {author}
  {\bibfnamefont {F.}~\bibnamefont {Dietrich}}, \ and\ \bibinfo {author}
  {\bibfnamefont {K.}~\bibnamefont {Droeder}},\ }\bibfield  {title} {\enquote
  {\bibinfo {title} {Current status and challenges for automotive battery
  production technologies},}\ }\href@noop {} {\bibfield  {journal} {\bibinfo
  {journal} {Nature Energy}\ }\textbf {\bibinfo {volume} {3}},\ \bibinfo
  {pages} {290--300} (\bibinfo {year} {2018})}\BibitemShut {NoStop}%
\bibitem [{\citenamefont {Cavazzoni}\ \emph {et~al.}(1999)\citenamefont
  {Cavazzoni}, \citenamefont {Chiarotti}, \citenamefont {Scandolo},
  \citenamefont {Tosatti}, \citenamefont {Bernasconi},\ and\ \citenamefont
  {Parrinello}}]{Cavazzoni1999}%
  \BibitemOpen
  \bibfield  {author} {\bibinfo {author} {\bibfnamefont {C.}~\bibnamefont
  {Cavazzoni}}, \bibinfo {author} {\bibfnamefont {G.}~\bibnamefont
  {Chiarotti}}, \bibinfo {author} {\bibfnamefont {S.}~\bibnamefont {Scandolo}},
  \bibinfo {author} {\bibfnamefont {E.}~\bibnamefont {Tosatti}}, \bibinfo
  {author} {\bibfnamefont {M.}~\bibnamefont {Bernasconi}}, \ and\ \bibinfo
  {author} {\bibfnamefont {M.}~\bibnamefont {Parrinello}},\ }\bibfield  {title}
  {\enquote {\bibinfo {title} {Superionic and metallic states of water and
  ammonia at giant planet conditions},}\ }\href@noop {} {\bibfield  {journal}
  {\bibinfo  {journal} {Science}\ }\textbf {\bibinfo {volume} {283}},\ \bibinfo
  {pages} {44--46} (\bibinfo {year} {1999})}\BibitemShut {NoStop}%
\bibitem [{\citenamefont {Redmer}\ \emph {et~al.}(2011)\citenamefont {Redmer},
  \citenamefont {Mattsson}, \citenamefont {Nettelmann},\ and\ \citenamefont
  {French}}]{Redmer2011}%
  \BibitemOpen
  \bibfield  {author} {\bibinfo {author} {\bibfnamefont {R.}~\bibnamefont
  {Redmer}}, \bibinfo {author} {\bibfnamefont {T.~R.}\ \bibnamefont
  {Mattsson}}, \bibinfo {author} {\bibfnamefont {N.}~\bibnamefont
  {Nettelmann}}, \ and\ \bibinfo {author} {\bibfnamefont {M.}~\bibnamefont
  {French}},\ }\bibfield  {title} {\enquote {\bibinfo {title} {The phase
  diagram of water and the magnetic fields of uranus and neptune},}\
  }\href@noop {} {\bibfield  {journal} {\bibinfo  {journal} {Icarus}\ }\textbf
  {\bibinfo {volume} {211}},\ \bibinfo {pages} {798--803} (\bibinfo {year}
  {2011})}\BibitemShut {NoStop}%
\bibitem [{\citenamefont {Nettelmann}\ \emph {et~al.}(2016)\citenamefont
  {Nettelmann}, \citenamefont {Wang}, \citenamefont {Fortney}, \citenamefont
  {Hamel}, \citenamefont {Yellamilli}, \citenamefont {Bethkenhagen},\ and\
  \citenamefont {Redmer}}]{Nettelmann2016}%
  \BibitemOpen
  \bibfield  {author} {\bibinfo {author} {\bibfnamefont {N.}~\bibnamefont
  {Nettelmann}}, \bibinfo {author} {\bibfnamefont {K.}~\bibnamefont {Wang}},
  \bibinfo {author} {\bibfnamefont {J.~J.}\ \bibnamefont {Fortney}}, \bibinfo
  {author} {\bibfnamefont {S.}~\bibnamefont {Hamel}}, \bibinfo {author}
  {\bibfnamefont {S.}~\bibnamefont {Yellamilli}}, \bibinfo {author}
  {\bibfnamefont {M.}~\bibnamefont {Bethkenhagen}}, \ and\ \bibinfo {author}
  {\bibfnamefont {R.}~\bibnamefont {Redmer}},\ }\bibfield  {title} {\enquote
  {\bibinfo {title} {Uranus evolution models with simple thermal boundary
  layers},}\ }\href@noop {} {\bibfield  {journal} {\bibinfo  {journal}
  {Icarus}\ }\textbf {\bibinfo {volume} {275}},\ \bibinfo {pages} {107--116}
  (\bibinfo {year} {2016})}\BibitemShut {NoStop}%
\bibitem [{\citenamefont {French}, \citenamefont {Hamel},\ and\ \citenamefont
  {Redmer}(2011)}]{French2011}%
  \BibitemOpen
  \bibfield  {author} {\bibinfo {author} {\bibfnamefont {M.}~\bibnamefont
  {French}}, \bibinfo {author} {\bibfnamefont {S.}~\bibnamefont {Hamel}}, \
  and\ \bibinfo {author} {\bibfnamefont {R.}~\bibnamefont {Redmer}},\
  }\bibfield  {title} {\enquote {\bibinfo {title} {Dynamical screening and
  ionic conductivity in water from ab initio simulations},}\ }\href
  {https://doi.org/10.1103/PhysRevLett.107.185901} {\bibfield  {journal}
  {\bibinfo  {journal} {Phys. Rev. Lett.}\ }\textbf {\bibinfo {volume} {107}},\
  \bibinfo {pages} {185901} (\bibinfo {year} {2011})}\BibitemShut {NoStop}%
\bibitem [{\citenamefont {Grasselli}, \citenamefont {Stixrude},\ and\
  \citenamefont {Baroni}(2020)}]{Grasselli2020}%
  \BibitemOpen
  \bibfield  {author} {\bibinfo {author} {\bibfnamefont {F.}~\bibnamefont
  {Grasselli}}, \bibinfo {author} {\bibfnamefont {L.}~\bibnamefont {Stixrude}},
  \ and\ \bibinfo {author} {\bibfnamefont {S.}~\bibnamefont {Baroni}},\
  }\bibfield  {title} {\enquote {\bibinfo {title} {Heat and charge transport in
  h 2 o at ice-giant conditions from ab initio molecular dynamics
  simulations},}\ }\href {https://doi.org/10.1038/s41467-020-17275-5}
  {\bibfield  {journal} {\bibinfo  {journal} {Nat. Commun.}\ }\textbf {\bibinfo
  {volume} {11}},\ \bibinfo {pages} {1--7} (\bibinfo {year}
  {2020})}\BibitemShut {NoStop}%
\bibitem [{\citenamefont {Millot}\ \emph {et~al.}(2018)\citenamefont {Millot},
  \citenamefont {Hamel}, \citenamefont {Rygg}, \citenamefont {Celliers},
  \citenamefont {Collins}, \citenamefont {Coppari}, \citenamefont
  {Fratanduono}, \citenamefont {Jeanloz}, \citenamefont {Swift},\ and\
  \citenamefont {Eggert}}]{Millot2018}%
  \BibitemOpen
  \bibfield  {author} {\bibinfo {author} {\bibfnamefont {M.}~\bibnamefont
  {Millot}}, \bibinfo {author} {\bibfnamefont {S.}~\bibnamefont {Hamel}},
  \bibinfo {author} {\bibfnamefont {J.~R.}\ \bibnamefont {Rygg}}, \bibinfo
  {author} {\bibfnamefont {P.~M.}\ \bibnamefont {Celliers}}, \bibinfo {author}
  {\bibfnamefont {G.~W.}\ \bibnamefont {Collins}}, \bibinfo {author}
  {\bibfnamefont {F.}~\bibnamefont {Coppari}}, \bibinfo {author} {\bibfnamefont
  {D.~E.}\ \bibnamefont {Fratanduono}}, \bibinfo {author} {\bibfnamefont
  {R.}~\bibnamefont {Jeanloz}}, \bibinfo {author} {\bibfnamefont {D.~C.}\
  \bibnamefont {Swift}}, \ and\ \bibinfo {author} {\bibfnamefont {J.~H.}\
  \bibnamefont {Eggert}},\ }\bibfield  {title} {\enquote {\bibinfo {title}
  {Experimental evidence for superionic water ice using shock compression},}\
  }\href@noop {} {\bibfield  {journal} {\bibinfo  {journal} {Nature Physics}\
  }\textbf {\bibinfo {volume} {14}},\ \bibinfo {pages} {297--302} (\bibinfo
  {year} {2018})}\BibitemShut {NoStop}%
\bibitem [{\citenamefont {Millot}\ \emph {et~al.}(2019)\citenamefont {Millot},
  \citenamefont {Coppari}, \citenamefont {Rygg}, \citenamefont {Barrios},
  \citenamefont {Hamel}, \citenamefont {Swift},\ and\ \citenamefont
  {Eggert}}]{Millot2019}%
  \BibitemOpen
  \bibfield  {author} {\bibinfo {author} {\bibfnamefont {M.}~\bibnamefont
  {Millot}}, \bibinfo {author} {\bibfnamefont {F.}~\bibnamefont {Coppari}},
  \bibinfo {author} {\bibfnamefont {J.~R.}\ \bibnamefont {Rygg}}, \bibinfo
  {author} {\bibfnamefont {A.~C.}\ \bibnamefont {Barrios}}, \bibinfo {author}
  {\bibfnamefont {S.}~\bibnamefont {Hamel}}, \bibinfo {author} {\bibfnamefont
  {D.~C.}\ \bibnamefont {Swift}}, \ and\ \bibinfo {author} {\bibfnamefont
  {J.~H.}\ \bibnamefont {Eggert}},\ }\bibfield  {title} {\enquote {\bibinfo
  {title} {Nanosecond x-ray diffraction of shock-compressed superionic water
  ice},}\ }\href@noop {} {\bibfield  {journal} {\bibinfo  {journal} {Nature}\
  }\textbf {\bibinfo {volume} {569}},\ \bibinfo {pages} {251--255} (\bibinfo
  {year} {2019})}\BibitemShut {NoStop}%
\bibitem [{\citenamefont {Podolak}, \citenamefont {Helled},\ and\ \citenamefont
  {Schubert}(2019)}]{Podolak2019}%
  \BibitemOpen
  \bibfield  {author} {\bibinfo {author} {\bibfnamefont {M.}~\bibnamefont
  {Podolak}}, \bibinfo {author} {\bibfnamefont {R.}~\bibnamefont {Helled}}, \
  and\ \bibinfo {author} {\bibfnamefont {G.}~\bibnamefont {Schubert}},\
  }\bibfield  {title} {\enquote {\bibinfo {title} {Effect of non-adiabatic
  thermal profiles on the inferred compositions of uranus and neptune},}\
  }\href@noop {} {\bibfield  {journal} {\bibinfo  {journal} {Monthly Notices of
  the Royal Astronomical Society}\ }\textbf {\bibinfo {volume} {487}},\
  \bibinfo {pages} {2653--2664} (\bibinfo {year} {2019})}\BibitemShut {NoStop}%
\bibitem [{\citenamefont {Stixrude}, \citenamefont {Baroni},\ and\
  \citenamefont {Grasselli}(2021)}]{Stixrude2021}%
  \BibitemOpen
  \bibfield  {author} {\bibinfo {author} {\bibfnamefont {L.}~\bibnamefont
  {Stixrude}}, \bibinfo {author} {\bibfnamefont {S.}~\bibnamefont {Baroni}}, \
  and\ \bibinfo {author} {\bibfnamefont {F.}~\bibnamefont {Grasselli}},\
  }\bibfield  {title} {\enquote {\bibinfo {title} {Thermal and tidal evolution
  of uranus with a growing frozen core},}\ }\href@noop {} {\bibfield  {journal}
  {\bibinfo  {journal} {The Planetary Science Journal}\ }\textbf {\bibinfo
  {volume} {2}},\ \bibinfo {pages} {222} (\bibinfo {year} {2021})}\BibitemShut
  {NoStop}%
\bibitem [{\citenamefont {Allen}\ and\ \citenamefont
  {Tildesley}(2017)}]{Allen2017}%
  \BibitemOpen
  \bibfield  {author} {\bibinfo {author} {\bibfnamefont {M.~P.}\ \bibnamefont
  {Allen}}\ and\ \bibinfo {author} {\bibfnamefont {D.~J.}\ \bibnamefont
  {Tildesley}},\ }\href@noop {} {\emph {\bibinfo {title} {Computer simulation
  of liquids}}}\ (\bibinfo  {publisher} {Oxford university press},\ \bibinfo
  {year} {2017})\BibitemShut {NoStop}%
\bibitem [{\citenamefont {Huang}\ \emph {et~al.}(2021)\citenamefont {Huang},
  \citenamefont {Zhang}, \citenamefont {Wang}, \citenamefont {Zhao},
  \citenamefont {Cheng},\ and\ \citenamefont {E}}]{Huang2021}%
  \BibitemOpen
  \bibfield  {author} {\bibinfo {author} {\bibfnamefont {J.}~\bibnamefont
  {Huang}}, \bibinfo {author} {\bibfnamefont {L.}~\bibnamefont {Zhang}},
  \bibinfo {author} {\bibfnamefont {H.}~\bibnamefont {Wang}}, \bibinfo {author}
  {\bibfnamefont {J.}~\bibnamefont {Zhao}}, \bibinfo {author} {\bibfnamefont
  {J.}~\bibnamefont {Cheng}}, \ and\ \bibinfo {author} {\bibfnamefont
  {W.}~\bibnamefont {E}},\ }\bibfield  {title} {\enquote {\bibinfo {title}
  {Deep potential generation scheme and simulation protocol for the
  li10gep2s12-type superionic conductors},}\ }\href@noop {} {\bibfield
  {journal} {\bibinfo  {journal} {The Journal of Chemical Physics}\ }\textbf
  {\bibinfo {volume} {154}},\ \bibinfo {pages} {094703} (\bibinfo {year}
  {2021})}\BibitemShut {NoStop}%
\bibitem [{\citenamefont {Musil}\ \emph {et~al.}(2021)\citenamefont {Musil},
  \citenamefont {Grisafi}, \citenamefont {Bart{\'o}k}, \citenamefont {Ortner},
  \citenamefont {Cs{\'a}nyi},\ and\ \citenamefont {Ceriotti}}]{Musil2021}%
  \BibitemOpen
  \bibfield  {author} {\bibinfo {author} {\bibfnamefont {F.}~\bibnamefont
  {Musil}}, \bibinfo {author} {\bibfnamefont {A.}~\bibnamefont {Grisafi}},
  \bibinfo {author} {\bibfnamefont {A.~P.}\ \bibnamefont {Bart{\'o}k}},
  \bibinfo {author} {\bibfnamefont {C.}~\bibnamefont {Ortner}}, \bibinfo
  {author} {\bibfnamefont {G.}~\bibnamefont {Cs{\'a}nyi}}, \ and\ \bibinfo
  {author} {\bibfnamefont {M.}~\bibnamefont {Ceriotti}},\ }\bibfield  {title}
  {\enquote {\bibinfo {title} {Physics-inspired structural representations for
  molecules and materials},}\ }\href@noop {} {\bibfield  {journal} {\bibinfo
  {journal} {Chemical Reviews}\ }\textbf {\bibinfo {volume} {121}},\ \bibinfo
  {pages} {9759--9815} (\bibinfo {year} {2021})}\BibitemShut {NoStop}%
\bibitem [{\citenamefont {Muy}\ \emph {et~al.}(2019)\citenamefont {Muy},
  \citenamefont {Voss}, \citenamefont {Schlem}, \citenamefont {Koerver},
  \citenamefont {Sedlmaier}, \citenamefont {Maglia}, \citenamefont {Lamp},
  \citenamefont {Zeier},\ and\ \citenamefont {Shao-Horn}}]{Muy2019}%
  \BibitemOpen
  \bibfield  {author} {\bibinfo {author} {\bibfnamefont {S.}~\bibnamefont
  {Muy}}, \bibinfo {author} {\bibfnamefont {J.}~\bibnamefont {Voss}}, \bibinfo
  {author} {\bibfnamefont {R.}~\bibnamefont {Schlem}}, \bibinfo {author}
  {\bibfnamefont {R.}~\bibnamefont {Koerver}}, \bibinfo {author} {\bibfnamefont
  {S.~J.}\ \bibnamefont {Sedlmaier}}, \bibinfo {author} {\bibfnamefont
  {F.}~\bibnamefont {Maglia}}, \bibinfo {author} {\bibfnamefont
  {P.}~\bibnamefont {Lamp}}, \bibinfo {author} {\bibfnamefont {W.~G.}\
  \bibnamefont {Zeier}}, \ and\ \bibinfo {author} {\bibfnamefont
  {Y.}~\bibnamefont {Shao-Horn}},\ }\bibfield  {title} {\enquote {\bibinfo
  {title} {High-throughput screening of solid-state {Li}-ion conductors using
  lattice-dynamics descriptors},}\ }\href@noop {} {\bibfield  {journal}
  {\bibinfo  {journal} {Iscience}\ }\textbf {\bibinfo {volume} {16}},\ \bibinfo
  {pages} {270--282} (\bibinfo {year} {2019})}\BibitemShut {NoStop}%
\bibitem [{\citenamefont {Kahle}, \citenamefont {Marcolongo},\ and\
  \citenamefont {Marzari}(2020)}]{Kahle2020}%
  \BibitemOpen
  \bibfield  {author} {\bibinfo {author} {\bibfnamefont {L.}~\bibnamefont
  {Kahle}}, \bibinfo {author} {\bibfnamefont {A.}~\bibnamefont {Marcolongo}}, \
  and\ \bibinfo {author} {\bibfnamefont {N.}~\bibnamefont {Marzari}},\
  }\bibfield  {title} {\enquote {\bibinfo {title} {High-throughput
  computational screening for solid-state {Li}-ion conductors},}\ }\href@noop {}
  {\bibfield  {journal} {\bibinfo  {journal} {Energy \& Environmental Science}\
  }\textbf {\bibinfo {volume} {13}},\ \bibinfo {pages} {928--948} (\bibinfo
  {year} {2020})}\BibitemShut {NoStop}%
\bibitem [{\citenamefont {Materzanini}\ \emph {et~al.}(2021)\citenamefont
  {Materzanini}, \citenamefont {Kahle}, \citenamefont {Marcolongo},\ and\
  \citenamefont {Marzari}}]{Materzanini2021}%
  \BibitemOpen
  \bibfield  {author} {\bibinfo {author} {\bibfnamefont {G.}~\bibnamefont
  {Materzanini}}, \bibinfo {author} {\bibfnamefont {L.}~\bibnamefont {Kahle}},
  \bibinfo {author} {\bibfnamefont {A.}~\bibnamefont {Marcolongo}}, \ and\
  \bibinfo {author} {\bibfnamefont {N.}~\bibnamefont {Marzari}},\ }\bibfield
  {title} {\enquote {\bibinfo {title} {High {Li}-ion conductivity in tetragonal
  lgpo: A comparative first-principles study against known LISICON and LGPS
  phases},}\ }\href@noop {} {\bibfield  {journal} {\bibinfo  {journal}
  {Physical Review Materials}\ }\textbf {\bibinfo {volume} {5}},\ \bibinfo
  {pages} {035408} (\bibinfo {year} {2021})}\BibitemShut {NoStop}%
\bibitem [{\citenamefont {Celebi}\ \emph {et~al.}(2021)\citenamefont {Celebi},
  \citenamefont {Jamali}, \citenamefont {Bardow}, \citenamefont {Vlugt},\ and\
  \citenamefont {Moultos}}]{Celebi2021}%
  \BibitemOpen
  \bibfield  {author} {\bibinfo {author} {\bibfnamefont {A.~T.}\ \bibnamefont
  {Celebi}}, \bibinfo {author} {\bibfnamefont {S.~H.}\ \bibnamefont {Jamali}},
  \bibinfo {author} {\bibfnamefont {A.}~\bibnamefont {Bardow}}, \bibinfo
  {author} {\bibfnamefont {T.~J.}\ \bibnamefont {Vlugt}}, \ and\ \bibinfo
  {author} {\bibfnamefont {O.~A.}\ \bibnamefont {Moultos}},\ }\bibfield
  {title} {\enquote {\bibinfo {title} {Finite-size effects of diffusion
  coefficients computed from molecular dynamics: a review of what we have
  learned so far},}\ }\href@noop {} {\bibfield  {journal} {\bibinfo  {journal}
  {Molecular Simulation}\ }\textbf {\bibinfo {volume} {47}},\ \bibinfo {pages}
  {831--845} (\bibinfo {year} {2021})}\BibitemShut {NoStop}%
\bibitem [{\citenamefont {D{\"u}nweg}\ and\ \citenamefont
  {Kremer}(1993)}]{Dunweg1993}%
  \BibitemOpen
  \bibfield  {author} {\bibinfo {author} {\bibfnamefont {B.}~\bibnamefont
  {D{\"u}nweg}}\ and\ \bibinfo {author} {\bibfnamefont {K.}~\bibnamefont
  {Kremer}},\ }\bibfield  {title} {\enquote {\bibinfo {title} {Molecular
  dynamics simulation of a polymer chain in solution},}\ }\href@noop {}
  {\bibfield  {journal} {\bibinfo  {journal} {The Journal of chemical physics}\
  }\textbf {\bibinfo {volume} {99}},\ \bibinfo {pages} {6983--6997} (\bibinfo
  {year} {1993})}\BibitemShut {NoStop}%
\bibitem [{\citenamefont {Yeh}\ and\ \citenamefont {Hummer}(2004)}]{Yeh2004}%
  \BibitemOpen
  \bibfield  {author} {\bibinfo {author} {\bibfnamefont {I.-C.}\ \bibnamefont
  {Yeh}}\ and\ \bibinfo {author} {\bibfnamefont {G.}~\bibnamefont {Hummer}},\
  }\bibfield  {title} {\enquote {\bibinfo {title} {System-size dependence of
  diffusion coefficients and viscosities from molecular dynamics simulations
  with periodic boundary conditions},}\ }\href@noop {} {\bibfield  {journal}
  {\bibinfo  {journal} {The Journal of Physical Chemistry B}\ }\textbf
  {\bibinfo {volume} {108}},\ \bibinfo {pages} {15873--15879} (\bibinfo {year}
  {2004})}\BibitemShut {NoStop}%
\bibitem [{\citenamefont {Moultos}\ \emph {et~al.}(2016)\citenamefont
  {Moultos}, \citenamefont {Zhang}, \citenamefont {Tsimpanogiannis},
  \citenamefont {Economou},\ and\ \citenamefont {Maginn}}]{Moultos2016}%
  \BibitemOpen
  \bibfield  {author} {\bibinfo {author} {\bibfnamefont {O.~A.}\ \bibnamefont
  {Moultos}}, \bibinfo {author} {\bibfnamefont {Y.}~\bibnamefont {Zhang}},
  \bibinfo {author} {\bibfnamefont {I.~N.}\ \bibnamefont {Tsimpanogiannis}},
  \bibinfo {author} {\bibfnamefont {I.~G.}\ \bibnamefont {Economou}}, \ and\
  \bibinfo {author} {\bibfnamefont {E.~J.}\ \bibnamefont {Maginn}},\ }\bibfield
   {title} {\enquote {\bibinfo {title} {System-size corrections for
  self-diffusion coefficients calculated from molecular dynamics simulations:
  The case of co2, n-alkanes, and poly (ethylene glycol) dimethyl ethers},}\
  }\href@noop {} {\bibfield  {journal} {\bibinfo  {journal} {The Journal of
  Chemical Physics}\ }\textbf {\bibinfo {volume} {145}},\ \bibinfo {pages}
  {074109} (\bibinfo {year} {2016})}\BibitemShut {NoStop}%
\bibitem [{\citenamefont {Jamali}\ \emph {et~al.}(2018)\citenamefont {Jamali},
  \citenamefont {Wolff}, \citenamefont {Becker}, \citenamefont {Bardow},
  \citenamefont {Vlugt},\ and\ \citenamefont {Moultos}}]{Jamali2018}%
  \BibitemOpen
  \bibfield  {author} {\bibinfo {author} {\bibfnamefont {S.~H.}\ \bibnamefont
  {Jamali}}, \bibinfo {author} {\bibfnamefont {L.}~\bibnamefont {Wolff}},
  \bibinfo {author} {\bibfnamefont {T.~M.}\ \bibnamefont {Becker}}, \bibinfo
  {author} {\bibfnamefont {A.}~\bibnamefont {Bardow}}, \bibinfo {author}
  {\bibfnamefont {T.~J.}\ \bibnamefont {Vlugt}}, \ and\ \bibinfo {author}
  {\bibfnamefont {O.~A.}\ \bibnamefont {Moultos}},\ }\bibfield  {title}
  {\enquote {\bibinfo {title} {Finite-size effects of binary mutual diffusion
  coefficients from molecular dynamics},}\ }\href@noop {} {\bibfield  {journal}
  {\bibinfo  {journal} {Journal of chemical theory and computation}\ }\textbf
  {\bibinfo {volume} {14}},\ \bibinfo {pages} {2667--2677} (\bibinfo {year}
  {2018})}\BibitemShut {NoStop}%
\bibitem [{\citenamefont {Jamali}\ \emph {et~al.}(2020)\citenamefont {Jamali},
  \citenamefont {Bardow}, \citenamefont {Vlugt},\ and\ \citenamefont
  {Moultos}}]{Jamali2020}%
  \BibitemOpen
  \bibfield  {author} {\bibinfo {author} {\bibfnamefont {S.~H.}\ \bibnamefont
  {Jamali}}, \bibinfo {author} {\bibfnamefont {A.}~\bibnamefont {Bardow}},
  \bibinfo {author} {\bibfnamefont {T.~J.}\ \bibnamefont {Vlugt}}, \ and\
  \bibinfo {author} {\bibfnamefont {O.~A.}\ \bibnamefont {Moultos}},\
  }\bibfield  {title} {\enquote {\bibinfo {title} {Generalized form for
  finite-size corrections in mutual diffusion coefficients of multicomponent
  mixtures obtained from equilibrium molecular dynamics simulation},}\
  }\href@noop {} {\bibfield  {journal} {\bibinfo  {journal} {Journal of
  chemical theory and computation}\ }\textbf {\bibinfo {volume} {16}},\
  \bibinfo {pages} {3799--3806} (\bibinfo {year} {2020})}\BibitemShut {NoStop}%
\bibitem [{\citenamefont {Shao}\ \emph {et~al.}(2020)\citenamefont {Shao},
  \citenamefont {Shigenobu}, \citenamefont {Watanabe},\ and\ \citenamefont
  {Zhang}}]{Shao2020}%
  \BibitemOpen
  \bibfield  {author} {\bibinfo {author} {\bibfnamefont {Y.}~\bibnamefont
  {Shao}}, \bibinfo {author} {\bibfnamefont {K.}~\bibnamefont {Shigenobu}},
  \bibinfo {author} {\bibfnamefont {M.}~\bibnamefont {Watanabe}}, \ and\
  \bibinfo {author} {\bibfnamefont {C.}~\bibnamefont {Zhang}},\ }\bibfield
  {title} {\enquote {\bibinfo {title} {Role of viscosity in deviations from the
  nernst--einstein relation},}\ }\href@noop {} {\bibfield  {journal} {\bibinfo
  {journal} {The Journal of Physical Chemistry B}\ }\textbf {\bibinfo {volume}
  {124}},\ \bibinfo {pages} {4774--4780} (\bibinfo {year} {2020})}\BibitemShut
  {NoStop}%
\bibitem [{\citenamefont {Zeyher}(1978)}]{Zeyher1978}%
  \BibitemOpen
  \bibfield  {author} {\bibinfo {author} {\bibfnamefont {R.}~\bibnamefont
  {Zeyher}},\ }\bibfield  {title} {\enquote {\bibinfo {title} {Hydrodynamics of
  superionic conductors},}\ }\href@noop {} {\bibfield  {journal} {\bibinfo
  {journal} {Zeitschrift f{\"u}r Physik B Condensed Matter}\ }\textbf {\bibinfo
  {volume} {31}},\ \bibinfo {pages} {127--142} (\bibinfo {year}
  {1978})}\BibitemShut {NoStop}%
\bibitem [{\citenamefont {Dieterich}, \citenamefont {Fulde},\ and\
  \citenamefont {Peschel}(1980)}]{Dieterich1980}%
  \BibitemOpen
  \bibfield  {author} {\bibinfo {author} {\bibfnamefont {W.}~\bibnamefont
  {Dieterich}}, \bibinfo {author} {\bibfnamefont {P.}~\bibnamefont {Fulde}}, \
  and\ \bibinfo {author} {\bibfnamefont {I.}~\bibnamefont {Peschel}},\
  }\bibfield  {title} {\enquote {\bibinfo {title} {Theoretical models for
  superionic conductors},}\ }\href@noop {} {\bibfield  {journal} {\bibinfo
  {journal} {Advances in Physics}\ }\textbf {\bibinfo {volume} {29}},\ \bibinfo
  {pages} {527--605} (\bibinfo {year} {1980})}\BibitemShut {NoStop}%
\bibitem [{\citenamefont {Dixon}\ and\ \citenamefont
  {Gillan}(1980{\natexlab{a}})}]{Dixon1980}%
  \BibitemOpen
  \bibfield  {author} {\bibinfo {author} {\bibfnamefont {M.}~\bibnamefont
  {Dixon}}\ and\ \bibinfo {author} {\bibfnamefont {M.}~\bibnamefont {Gillan}},\
  }\bibfield  {title} {\enquote {\bibinfo {title} {Computer simulation of fast
  ion transport in fluorites},}\ }\href@noop {} {\bibfield  {journal} {\bibinfo
   {journal} {Le Journal de Physique Colloques}\ }\textbf {\bibinfo {volume}
  {41}},\ \bibinfo {pages} {C6--24} (\bibinfo {year}
  {1980}{\natexlab{a}})}\BibitemShut {NoStop}%
\bibitem [{Note1()}]{Note1}%
  \BibitemOpen
  \bibinfo {note} {Notice that, even in the case of liquids, the specific
  functional dependence of the thermal conductivity on the system size seems to
  be qualitatively affected by the particular pressure and temperature
  conditions of the simulation \cite {Puligheddu2020}}\BibitemShut {NoStop}%
\bibitem [{\citenamefont {Botan}, \citenamefont {Marry},\ and\ \citenamefont
  {Rotenberg}(2015)}]{Botan2015}%
  \BibitemOpen
  \bibfield  {author} {\bibinfo {author} {\bibfnamefont {A.}~\bibnamefont
  {Botan}}, \bibinfo {author} {\bibfnamefont {V.}~\bibnamefont {Marry}}, \ and\
  \bibinfo {author} {\bibfnamefont {B.}~\bibnamefont {Rotenberg}},\ }\bibfield
  {title} {\enquote {\bibinfo {title} {Diffusion in bulk liquids: finite-size
  effects in anisotropic systems},}\ }\href@noop {} {\bibfield  {journal}
  {\bibinfo  {journal} {Molecular Physics}\ }\textbf {\bibinfo {volume}
  {113}},\ \bibinfo {pages} {2674--2679} (\bibinfo {year} {2015})}\BibitemShut
  {NoStop}%
\bibitem [{\citenamefont {Dixon}\ and\ \citenamefont
  {Gillan}(1980{\natexlab{b}})}]{Dixon1980b}%
  \BibitemOpen
  \bibfield  {author} {\bibinfo {author} {\bibfnamefont {M.}~\bibnamefont
  {Dixon}}\ and\ \bibinfo {author} {\bibfnamefont {M.}~\bibnamefont {Gillan}},\
  }\bibfield  {title} {\enquote {\bibinfo {title} {Molecular dynamics
  simulation of fast-ion conduction in srcl2. ii. distribution of ions and
  specific heat anomaly},}\ }\href@noop {} {\bibfield  {journal} {\bibinfo
  {journal} {Journal of Physics C: Solid State Physics}\ }\textbf {\bibinfo
  {volume} {13}},\ \bibinfo {pages} {1919} (\bibinfo {year}
  {1980}{\natexlab{b}})}\BibitemShut {NoStop}%
\bibitem [{\citenamefont {Mohn}\ \emph {et~al.}(2021)\citenamefont {Mohn},
  \citenamefont {Krynski}, \citenamefont {Kob},\ and\ \citenamefont
  {Allan}}]{Mohn2021}%
  \BibitemOpen
  \bibfield  {author} {\bibinfo {author} {\bibfnamefont {C.~E.}\ \bibnamefont
  {Mohn}}, \bibinfo {author} {\bibfnamefont {M.}~\bibnamefont {Krynski}},
  \bibinfo {author} {\bibfnamefont {W.}~\bibnamefont {Kob}}, \ and\ \bibinfo
  {author} {\bibfnamefont {N.~L.}\ \bibnamefont {Allan}},\ }\bibfield  {title}
  {\enquote {\bibinfo {title} {Cooperative excitations in superionic pbf2},}\
  }\href@noop {} {\bibfield  {journal} {\bibinfo  {journal} {Philosophical
  Transactions of the Royal Society A}\ }\textbf {\bibinfo {volume} {379}},\
  \bibinfo {pages} {20190455} (\bibinfo {year} {2021})}\BibitemShut {NoStop}%
\bibitem [{Note2()}]{Note2}%
  \BibitemOpen
  \bibinfo {note} {Below the SI transition, the transient hopping mechanism is
  dominated by vacancy motion, while, in the SI phase, it can be \protect
  \textit {``attributed in roughly equal measure to vacancy and interstitial
  motion''} (\protect \textit {verbatim} from Ref.~\protect \rev@citealpnum
  {Gillan1980}).}\BibitemShut {Stop}%
\bibitem [{\citenamefont {Binder}(1987)}]{Binder1987}%
  \BibitemOpen
  \bibfield  {author} {\bibinfo {author} {\bibfnamefont {K.}~\bibnamefont
  {Binder}},\ }\bibfield  {title} {\enquote {\bibinfo {title} {Finite size
  effects on phase transitions},}\ }\href@noop {} {\bibfield  {journal}
  {\bibinfo  {journal} {Ferroelectrics}\ }\textbf {\bibinfo {volume} {73}},\
  \bibinfo {pages} {43--67} (\bibinfo {year} {1987})}\BibitemShut {NoStop}%
\bibitem [{\citenamefont {Binner}\ \emph {et~al.}(2006)\citenamefont {Binner},
  \citenamefont {Price}, \citenamefont {Reading},\ and\ \citenamefont
  {Vaidhyanathan}}]{Binner2006}%
  \BibitemOpen
  \bibfield  {author} {\bibinfo {author} {\bibfnamefont {J.}~\bibnamefont
  {Binner}}, \bibinfo {author} {\bibfnamefont {D.}~\bibnamefont {Price}},
  \bibinfo {author} {\bibfnamefont {M.}~\bibnamefont {Reading}}, \ and\
  \bibinfo {author} {\bibfnamefont {B.}~\bibnamefont {Vaidhyanathan}},\
  }\bibfield  {title} {\enquote {\bibinfo {title} {Modulated temperature
  calorimetry of silver iodide in the presence of microwave radiation},}\
  }\href@noop {} {\bibfield  {journal} {\bibinfo  {journal} {Thermochimica
  acta}\ }\textbf {\bibinfo {volume} {446}},\ \bibinfo {pages} {156--160}
  (\bibinfo {year} {2006})}\BibitemShut {NoStop}%
\bibitem [{\citenamefont {Fossati}, \citenamefont {Chartier},\ and\
  \citenamefont {Boulle}(2021)}]{Fossati2021}%
  \BibitemOpen
  \bibfield  {author} {\bibinfo {author} {\bibfnamefont {P.}~\bibnamefont
  {Fossati}}, \bibinfo {author} {\bibfnamefont {A.}~\bibnamefont {Chartier}}, \
  and\ \bibinfo {author} {\bibfnamefont {A.}~\bibnamefont {Boulle}},\
  }\bibfield  {title} {\enquote {\bibinfo {title} {Structural aspects of the
  superionic transition in ax2 compounds with the fluorite structure},}\
  }\href@noop {} {\bibfield  {journal} {\bibinfo  {journal} {Frontiers in
  chemistry}\ ,\ \bibinfo {pages} {746}} (\bibinfo {year} {2021})}\BibitemShut
  {NoStop}%
\bibitem [{\citenamefont {Walker}, \citenamefont {Dixon},\ and\ \citenamefont
  {Gillan}(1982)}]{Walker1982}%
  \BibitemOpen
  \bibfield  {author} {\bibinfo {author} {\bibfnamefont {A.}~\bibnamefont
  {Walker}}, \bibinfo {author} {\bibfnamefont {M.}~\bibnamefont {Dixon}}, \
  and\ \bibinfo {author} {\bibfnamefont {M.}~\bibnamefont {Gillan}},\
  }\bibfield  {title} {\enquote {\bibinfo {title} {Computer simulation of ionic
  disorder in high-temperature pbf2},}\ }\href@noop {} {\bibfield  {journal}
  {\bibinfo  {journal} {Journal of Physics C: Solid State Physics}\ }\textbf
  {\bibinfo {volume} {15}},\ \bibinfo {pages} {4061} (\bibinfo {year}
  {1982})}\BibitemShut {NoStop}%
\bibitem [{\citenamefont {Cooper}, \citenamefont {Rushton},\ and\ \citenamefont
  {Grimes}(2014)}]{Cooper2014a}%
  \BibitemOpen
  \bibfield  {author} {\bibinfo {author} {\bibfnamefont {M.}~\bibnamefont
  {Cooper}}, \bibinfo {author} {\bibfnamefont {M.}~\bibnamefont {Rushton}}, \
  and\ \bibinfo {author} {\bibfnamefont {R.}~\bibnamefont {Grimes}},\
  }\bibfield  {title} {\enquote {\bibinfo {title} {A many-body potential
  approach to modelling the thermomechanical properties of actinide oxides},}\
  }\href@noop {} {\bibfield  {journal} {\bibinfo  {journal} {Journal of
  Physics: Condensed Matter}\ }\textbf {\bibinfo {volume} {26}},\ \bibinfo
  {pages} {105401} (\bibinfo {year} {2014})}\BibitemShut {NoStop}%
\bibitem [{\citenamefont {Cooper}\ \emph {et~al.}(2014)\citenamefont {Cooper},
  \citenamefont {Murphy}, \citenamefont {Fossati}, \citenamefont {Rushton},\
  and\ \citenamefont {Grimes}}]{Cooper2014b}%
  \BibitemOpen
  \bibfield  {author} {\bibinfo {author} {\bibfnamefont {M.~W.}\ \bibnamefont
  {Cooper}}, \bibinfo {author} {\bibfnamefont {S.~T.}\ \bibnamefont {Murphy}},
  \bibinfo {author} {\bibfnamefont {P.~C.}\ \bibnamefont {Fossati}}, \bibinfo
  {author} {\bibfnamefont {M.~J.}\ \bibnamefont {Rushton}}, \ and\ \bibinfo
  {author} {\bibfnamefont {R.~W.}\ \bibnamefont {Grimes}},\ }\bibfield  {title}
  {\enquote {\bibinfo {title} {Thermophysical and anion diffusion properties of
  (u x, th1- x) o2},}\ }\href@noop {} {\bibfield  {journal} {\bibinfo
  {journal} {Proceedings of the Royal Society A: Mathematical, Physical and
  Engineering Sciences}\ }\textbf {\bibinfo {volume} {470}},\ \bibinfo {pages}
  {20140427} (\bibinfo {year} {2014})}\BibitemShut {NoStop}%
\bibitem [{\citenamefont {Niu}\ \emph {et~al.}(2018)\citenamefont {Niu},
  \citenamefont {Jing}, \citenamefont {Sun},\ and\ \citenamefont
  {Aluru}}]{Niu2018}%
  \BibitemOpen
  \bibfield  {author} {\bibinfo {author} {\bibfnamefont {H.}~\bibnamefont
  {Niu}}, \bibinfo {author} {\bibfnamefont {Y.}~\bibnamefont {Jing}}, \bibinfo
  {author} {\bibfnamefont {Y.}~\bibnamefont {Sun}}, \ and\ \bibinfo {author}
  {\bibfnamefont {N.~R.}\ \bibnamefont {Aluru}},\ }\bibfield  {title} {\enquote
  {\bibinfo {title} {Ab initio based interionic potential for silver iodide},}\
  }\href@noop {} {\bibfield  {journal} {\bibinfo  {journal} {Solid State
  Ionics}\ }\textbf {\bibinfo {volume} {325}},\ \bibinfo {pages} {102--111}
  (\bibinfo {year} {2018})}\BibitemShut {NoStop}%
\bibitem [{\citenamefont {Kvist}\ and\ \citenamefont
  {T{\"a}rneberg}(1970)}]{Kvist1970}%
  \BibitemOpen
  \bibfield  {author} {\bibinfo {author} {\bibfnamefont {A.}~\bibnamefont
  {Kvist}}\ and\ \bibinfo {author} {\bibfnamefont {R.}~\bibnamefont
  {T{\"a}rneberg}},\ }\bibfield  {title} {\enquote {\bibinfo {title}
  {Self-diffusion of silver ions in the cubic high temperature modification of
  silver iodide},}\ }\href@noop {} {\bibfield  {journal} {\bibinfo  {journal}
  {Zeitschrift f{\"u}r Naturforschung A}\ }\textbf {\bibinfo {volume} {25}},\
  \bibinfo {pages} {257--259} (\bibinfo {year} {1970})}\BibitemShut {NoStop}%
\bibitem [{\citenamefont {Parrinello}, \citenamefont {Rahman},\ and\
  \citenamefont {Vashishta}(1983)}]{Parrinello1983}%
  \BibitemOpen
  \bibfield  {author} {\bibinfo {author} {\bibfnamefont {M.}~\bibnamefont
  {Parrinello}}, \bibinfo {author} {\bibfnamefont {A.}~\bibnamefont {Rahman}},
  \ and\ \bibinfo {author} {\bibfnamefont {P.}~\bibnamefont {Vashishta}},\
  }\bibfield  {title} {\enquote {\bibinfo {title} {Structural transitions in
  superionic conductors},}\ }\href@noop {} {\bibfield  {journal} {\bibinfo
  {journal} {Physical review letters}\ }\textbf {\bibinfo {volume} {50}},\
  \bibinfo {pages} {1073} (\bibinfo {year} {1983})}\BibitemShut {NoStop}%
\bibitem [{\citenamefont {Thompson}\ \emph {et~al.}(2022)\citenamefont
  {Thompson}, \citenamefont {Aktulga}, \citenamefont {Berger}, \citenamefont
  {Bolintineanu}, \citenamefont {Brown}, \citenamefont {Crozier}, \citenamefont
  {in~'t Veld}, \citenamefont {Kohlmeyer}, \citenamefont {Moore}, \citenamefont
  {Nguyen}, \citenamefont {Shan}, \citenamefont {Stevens}, \citenamefont
  {Tranchida}, \citenamefont {Trott},\ and\ \citenamefont {Plimpton}}]{LAMMPS}%
  \BibitemOpen
  \bibfield  {author} {\bibinfo {author} {\bibfnamefont {A.~P.}\ \bibnamefont
  {Thompson}}, \bibinfo {author} {\bibfnamefont {H.~M.}\ \bibnamefont
  {Aktulga}}, \bibinfo {author} {\bibfnamefont {R.}~\bibnamefont {Berger}},
  \bibinfo {author} {\bibfnamefont {D.~S.}\ \bibnamefont {Bolintineanu}},
  \bibinfo {author} {\bibfnamefont {W.~M.}\ \bibnamefont {Brown}}, \bibinfo
  {author} {\bibfnamefont {P.~S.}\ \bibnamefont {Crozier}}, \bibinfo {author}
  {\bibfnamefont {P.~J.}\ \bibnamefont {in~'t Veld}}, \bibinfo {author}
  {\bibfnamefont {A.}~\bibnamefont {Kohlmeyer}}, \bibinfo {author}
  {\bibfnamefont {S.~G.}\ \bibnamefont {Moore}}, \bibinfo {author}
  {\bibfnamefont {T.~D.}\ \bibnamefont {Nguyen}}, \bibinfo {author}
  {\bibfnamefont {R.}~\bibnamefont {Shan}}, \bibinfo {author} {\bibfnamefont
  {M.~J.}\ \bibnamefont {Stevens}}, \bibinfo {author} {\bibfnamefont
  {J.}~\bibnamefont {Tranchida}}, \bibinfo {author} {\bibfnamefont
  {C.}~\bibnamefont {Trott}}, \ and\ \bibinfo {author} {\bibfnamefont {S.~J.}\
  \bibnamefont {Plimpton}},\ }\bibfield  {title} {\enquote {\bibinfo {title}
  {{LAMMPS} - a flexible simulation tool for particle-based materials modeling
  at the atomic, meso, and continuum scales},}\ }\href {\doibase
  10.1016/j.cpc.2021.108171} {\bibfield  {journal} {\bibinfo  {journal} {Comp.
  Phys. Comm.}\ }\textbf {\bibinfo {volume} {271}},\ \bibinfo {pages} {108171}
  (\bibinfo {year} {2022})}\BibitemShut {NoStop}%
\bibitem [{\citenamefont {Hockney}\ and\ \citenamefont
  {Eastwood}(2021)}]{Hockney2021}%
  \BibitemOpen
  \bibfield  {author} {\bibinfo {author} {\bibfnamefont {R.~W.}\ \bibnamefont
  {Hockney}}\ and\ \bibinfo {author} {\bibfnamefont {J.~W.}\ \bibnamefont
  {Eastwood}},\ }\href@noop {} {\emph {\bibinfo {title} {Computer simulation
  using particles}}}\ (\bibinfo  {publisher} {crc Press},\ \bibinfo {year}
  {2021})\BibitemShut {NoStop}%
\bibitem [{\citenamefont {Bussi}, \citenamefont {Donadio},\ and\ \citenamefont
  {Parrinello}(2007)}]{Bussi2007}%
  \BibitemOpen
  \bibfield  {author} {\bibinfo {author} {\bibfnamefont {G.}~\bibnamefont
  {Bussi}}, \bibinfo {author} {\bibfnamefont {D.}~\bibnamefont {Donadio}}, \
  and\ \bibinfo {author} {\bibfnamefont {M.}~\bibnamefont {Parrinello}},\
  }\bibfield  {title} {\enquote {\bibinfo {title} {Canonical sampling through
  velocity rescaling},}\ }\href@noop {} {\bibfield  {journal} {\bibinfo
  {journal} {The Journal of chemical physics}\ }\textbf {\bibinfo {volume}
  {126}},\ \bibinfo {pages} {014101} (\bibinfo {year} {2007})}\BibitemShut
  {NoStop}%
\bibitem [{\citenamefont {Naylor}(1945)}]{Naylor1945}%
  \BibitemOpen
  \bibfield  {author} {\bibinfo {author} {\bibfnamefont {B.}~\bibnamefont
  {Naylor}},\ }\bibfield  {title} {\enquote {\bibinfo {title} {Heat contents at
  high temperatures of magnesium and calcium fluorides1},}\ }\href@noop {}
  {\bibfield  {journal} {\bibinfo  {journal} {Journal of the American Chemical
  Society}\ }\textbf {\bibinfo {volume} {67}},\ \bibinfo {pages} {150--152}
  (\bibinfo {year} {1945})}\BibitemShut {NoStop}%
\bibitem [{\citenamefont {Derrington}, \citenamefont {Navrotsky},\ and\
  \citenamefont {O'Keeffe}(1976)}]{Derrington1976}%
  \BibitemOpen
  \bibfield  {author} {\bibinfo {author} {\bibfnamefont {C.}~\bibnamefont
  {Derrington}}, \bibinfo {author} {\bibfnamefont {A.}~\bibnamefont
  {Navrotsky}}, \ and\ \bibinfo {author} {\bibfnamefont {M.}~\bibnamefont
  {O'Keeffe}},\ }\bibfield  {title} {\enquote {\bibinfo {title} {High
  temperature heat content and diffuse transition of lead fluoride},}\
  }\href@noop {} {\bibfield  {journal} {\bibinfo  {journal} {Solid State
  Communications}\ }\textbf {\bibinfo {volume} {18}},\ \bibinfo {pages}
  {47--49} (\bibinfo {year} {1976})}\BibitemShut {NoStop}%
\bibitem [{\citenamefont {Ferdinand}\ and\ \citenamefont
  {Fisher}(1969)}]{Ferdinand1969}%
  \BibitemOpen
  \bibfield  {author} {\bibinfo {author} {\bibfnamefont {A.~E.}\ \bibnamefont
  {Ferdinand}}\ and\ \bibinfo {author} {\bibfnamefont {M.~E.}\ \bibnamefont
  {Fisher}},\ }\bibfield  {title} {\enquote {\bibinfo {title} {Bounded and
  inhomogeneous ising models. i. specific-heat anomaly of a finite lattice},}\
  }\href@noop {} {\bibfield  {journal} {\bibinfo  {journal} {Physical Review}\
  }\textbf {\bibinfo {volume} {185}},\ \bibinfo {pages} {832} (\bibinfo {year}
  {1969})}\BibitemShut {NoStop}%
\bibitem [{\citenamefont {Yakub}, \citenamefont {Ronchi},\ and\ \citenamefont
  {Staicu}(2007)}]{Yakub2007}%
  \BibitemOpen
  \bibfield  {author} {\bibinfo {author} {\bibfnamefont {E.}~\bibnamefont
  {Yakub}}, \bibinfo {author} {\bibfnamefont {C.}~\bibnamefont {Ronchi}}, \
  and\ \bibinfo {author} {\bibfnamefont {D.}~\bibnamefont {Staicu}},\
  }\bibfield  {title} {\enquote {\bibinfo {title} {Molecular dynamics
  simulation of premelting and melting phase transitions in stoichiometric
  uranium dioxide},}\ }\href@noop {} {\bibfield  {journal} {\bibinfo  {journal}
  {The Journal of chemical physics}\ }\textbf {\bibinfo {volume} {127}},\
  \bibinfo {pages} {094508} (\bibinfo {year} {2007})}\BibitemShut {NoStop}%
\bibitem [{\citenamefont {Fink}(2000)}]{Fink2000}%
  \BibitemOpen
  \bibfield  {author} {\bibinfo {author} {\bibfnamefont {J.}~\bibnamefont
  {Fink}},\ }\bibfield  {title} {\enquote {\bibinfo {title} {Thermophysical
  properties of uranium dioxide},}\ }\href@noop {} {\bibfield  {journal}
  {\bibinfo  {journal} {Journal of nuclear materials}\ }\textbf {\bibinfo
  {volume} {279}},\ \bibinfo {pages} {1--18} (\bibinfo {year}
  {2000})}\BibitemShut {NoStop}%
\bibitem [{Sup()}]{SupplMat}%
  \BibitemOpen
  \href@noop {} {}\bibinfo {note} {See Supplementary Material at [URL will be
  inserted by publisher], which also contains additional
  Refs.~\onlinecite{analisi,travis1,travis2,Fan2015,Ercole2017,Boone2019,Surblys2019,Simmons1971}.}\BibitemShut
  {Stop}%
\bibitem [{\citenamefont {Potashnikov}\ \emph {et~al.}(2013)\citenamefont
  {Potashnikov}, \citenamefont {Boyarchenkov}, \citenamefont {Nekrasov},\ and\
  \citenamefont {Kupryazhkin}}]{Potashnikov2013}%
  \BibitemOpen
  \bibfield  {author} {\bibinfo {author} {\bibfnamefont {S.}~\bibnamefont
  {Potashnikov}}, \bibinfo {author} {\bibfnamefont {A.}~\bibnamefont
  {Boyarchenkov}}, \bibinfo {author} {\bibfnamefont {K.}~\bibnamefont
  {Nekrasov}}, \ and\ \bibinfo {author} {\bibfnamefont {A.~Y.}\ \bibnamefont
  {Kupryazhkin}},\ }\bibfield  {title} {\enquote {\bibinfo {title}
  {High-precision molecular dynamics simulation of {UO$_2$}--{PuO$_2$}: Anion
  self-diffusion in {UO$_2$}},}\ }\href@noop {} {\bibfield  {journal} {\bibinfo
  {journal} {Journal of nuclear materials}\ }\textbf {\bibinfo {volume}
  {433}},\ \bibinfo {pages} {215--226} (\bibinfo {year} {2013})}\BibitemShut
  {NoStop}%
\bibitem [{\citenamefont {Kahle}, \citenamefont {Marcolongo},\ and\
  \citenamefont {Marzari}(2018)}]{Kahle2018}%
  \BibitemOpen
  \bibfield  {author} {\bibinfo {author} {\bibfnamefont {L.}~\bibnamefont
  {Kahle}}, \bibinfo {author} {\bibfnamefont {A.}~\bibnamefont {Marcolongo}}, \
  and\ \bibinfo {author} {\bibfnamefont {N.}~\bibnamefont {Marzari}},\
  }\bibfield  {title} {\enquote {\bibinfo {title} {Modeling lithium-ion
  solid-state electrolytes with a pinball model},}\ }\href@noop {} {\bibfield
  {journal} {\bibinfo  {journal} {Physical Review Materials}\ }\textbf
  {\bibinfo {volume} {2}},\ \bibinfo {pages} {065405} (\bibinfo {year}
  {2018})}\BibitemShut {NoStop}%
\bibitem [{\citenamefont {Fennell}\ and\ \citenamefont
  {Gezelter}(2006)}]{Fennell2006}%
  \BibitemOpen
  \bibfield  {author} {\bibinfo {author} {\bibfnamefont {C.~J.}\ \bibnamefont
  {Fennell}}\ and\ \bibinfo {author} {\bibfnamefont {J.~D.}\ \bibnamefont
  {Gezelter}},\ }\bibfield  {title} {\enquote {\bibinfo {title} {Is the ewald
  summation still necessary? pairwise alternatives to the accepted standard for
  long-range electrostatics},}\ }\href@noop {} {\bibfield  {journal} {\bibinfo
  {journal} {The Journal of chemical physics}\ }\textbf {\bibinfo {volume}
  {124}},\ \bibinfo {pages} {234104} (\bibinfo {year} {2006})}\BibitemShut
  {NoStop}%
\bibitem [{\citenamefont {Gillan}\ and\ \citenamefont
  {Dixon}(1980)}]{Gillan1980}%
  \BibitemOpen
  \bibfield  {author} {\bibinfo {author} {\bibfnamefont {M.}~\bibnamefont
  {Gillan}}\ and\ \bibinfo {author} {\bibfnamefont {M.}~\bibnamefont {Dixon}},\
  }\bibfield  {title} {\enquote {\bibinfo {title} {Molecular dynamics
  simulation of fast-ion conduction in srcl2. i. self-diffusion},}\ }\href@noop
  {} {\bibfield  {journal} {\bibinfo  {journal} {Journal of Physics C: Solid
  State Physics}\ }\textbf {\bibinfo {volume} {13}},\ \bibinfo {pages} {1901}
  (\bibinfo {year} {1980})}\BibitemShut {NoStop}%
\bibitem [{\citenamefont {Simoncelli}, \citenamefont {Marzari},\ and\
  \citenamefont {Mauri}(2019)}]{Simoncelli2019}%
  \BibitemOpen
  \bibfield  {author} {\bibinfo {author} {\bibfnamefont {M.}~\bibnamefont
  {Simoncelli}}, \bibinfo {author} {\bibfnamefont {N.}~\bibnamefont {Marzari}},
  \ and\ \bibinfo {author} {\bibfnamefont {F.}~\bibnamefont {Mauri}},\
  }\bibfield  {title} {\enquote {\bibinfo {title} {Unified theory of thermal
  transport in crystals and glasses},}\ }\href@noop {} {\bibfield  {journal}
  {\bibinfo  {journal} {Nature Physics}\ }\textbf {\bibinfo {volume} {15}},\
  \bibinfo {pages} {809--813} (\bibinfo {year} {2019})}\BibitemShut {NoStop}%
\bibitem [{\citenamefont {Isaeva}\ \emph {et~al.}(2019)\citenamefont {Isaeva},
  \citenamefont {Barbalinardo}, \citenamefont {Donadio},\ and\ \citenamefont
  {Baroni}}]{Isaeva2019}%
  \BibitemOpen
  \bibfield  {author} {\bibinfo {author} {\bibfnamefont {L.}~\bibnamefont
  {Isaeva}}, \bibinfo {author} {\bibfnamefont {G.}~\bibnamefont
  {Barbalinardo}}, \bibinfo {author} {\bibfnamefont {D.}~\bibnamefont
  {Donadio}}, \ and\ \bibinfo {author} {\bibfnamefont {S.}~\bibnamefont
  {Baroni}},\ }\bibfield  {title} {\enquote {\bibinfo {title} {Modeling heat
  transport in crystals and glasses from a unified lattice-dynamical
  approach},}\ }\href@noop {} {\bibfield  {journal} {\bibinfo  {journal}
  {Nature communications}\ }\textbf {\bibinfo {volume} {10}},\ \bibinfo {pages}
  {1--6} (\bibinfo {year} {2019})}\BibitemShut {NoStop}%
\bibitem [{\citenamefont {Allen}\ and\ \citenamefont
  {Feldman}(1993)}]{AllenFeldman}%
  \BibitemOpen
  \bibfield  {author} {\bibinfo {author} {\bibfnamefont {P.~B.}\ \bibnamefont
  {Allen}}\ and\ \bibinfo {author} {\bibfnamefont {J.~L.}\ \bibnamefont
  {Feldman}},\ }\bibfield  {title} {\enquote {\bibinfo {title} {Thermal
  conductivity of disordered harmonic solids},}\ }\href@noop {} {\bibfield
  {journal} {\bibinfo  {journal} {Physical Review B}\ }\textbf {\bibinfo
  {volume} {48}},\ \bibinfo {pages} {12581} (\bibinfo {year}
  {1993})}\BibitemShut {NoStop}%
\bibitem [{\citenamefont {Pegolo}, \citenamefont {Baroni},\ and\ \citenamefont
  {Grasselli}(2022)}]{Pegolo2021}%
  \BibitemOpen
  \bibfield  {author} {\bibinfo {author} {\bibfnamefont {P.}~\bibnamefont
  {Pegolo}}, \bibinfo {author} {\bibfnamefont {S.}~\bibnamefont {Baroni}}, \
  and\ \bibinfo {author} {\bibfnamefont {F.}~\bibnamefont {Grasselli}},\
  }\bibfield  {title} {\enquote {\bibinfo {title} {Temperature-and
  vacancy-concentration-dependence of heat transport in {Li$_3$ClO} from
  multi-method numerical simulations},}\ }\href@noop {} {\bibfield  {journal}
  {\bibinfo  {journal} {npj Computational Materials}\ }\textbf {\bibinfo
  {volume} {8}},\ \bibinfo {pages} {1--9} (\bibinfo {year} {2022})}\BibitemShut
  {NoStop}%
\bibitem [{Note3()}]{Note3}%
  \BibitemOpen
  \bibinfo {note} {This is also confirmed by the temperature, lower for smaller
  systems, at which the multicomponent analysis departs from the
  single-component one, which assumes no atomic diffusion (see Fig.~\ref
  {fig:kappa_multi_single}).}\BibitemShut {Stop}%
\bibitem [{\citenamefont {Popov}\ \emph {et~al.}(2017)\citenamefont {Popov},
  \citenamefont {Sidorov}, \citenamefont {Kul'chenkov}, \citenamefont
  {Anishchenko}, \citenamefont {Avetissov}, \citenamefont {Sorokin},\ and\
  \citenamefont {Fedorov}}]{Popov2017}%
  \BibitemOpen
  \bibfield  {author} {\bibinfo {author} {\bibfnamefont {P.}~\bibnamefont
  {Popov}}, \bibinfo {author} {\bibfnamefont {A.}~\bibnamefont {Sidorov}},
  \bibinfo {author} {\bibfnamefont {E.}~\bibnamefont {Kul'chenkov}}, \bibinfo
  {author} {\bibfnamefont {A.}~\bibnamefont {Anishchenko}}, \bibinfo {author}
  {\bibfnamefont {I.~C.}\ \bibnamefont {Avetissov}}, \bibinfo {author}
  {\bibfnamefont {N.}~\bibnamefont {Sorokin}}, \ and\ \bibinfo {author}
  {\bibfnamefont {P.}~\bibnamefont {Fedorov}},\ }\bibfield  {title} {\enquote
  {\bibinfo {title} {Thermal conductivity and expansion of {PbF$_2$} single
  crystals},}\ }\href@noop {} {\bibfield  {journal} {\bibinfo  {journal}
  {Ionics}\ }\textbf {\bibinfo {volume} {23}},\ \bibinfo {pages} {233--239}
  (\bibinfo {year} {2017})}\BibitemShut {NoStop}%
\bibitem [{\citenamefont {Lindan}\ and\ \citenamefont
  {Gillan}(1991)}]{Lindan1991}%
  \BibitemOpen
  \bibfield  {author} {\bibinfo {author} {\bibfnamefont {P.}~\bibnamefont
  {Lindan}}\ and\ \bibinfo {author} {\bibfnamefont {M.}~\bibnamefont
  {Gillan}},\ }\bibfield  {title} {\enquote {\bibinfo {title} {A molecular
  dynamics study of the thermal conductivity of caf2 and uo2},}\ }\href@noop {}
  {\bibfield  {journal} {\bibinfo  {journal} {Journal of Physics: Condensed
  Matter}\ }\textbf {\bibinfo {volume} {3}},\ \bibinfo {pages} {3929} (\bibinfo
  {year} {1991})}\BibitemShut {NoStop}%
\bibitem [{\citenamefont {Goetz}\ and\ \citenamefont
  {Cowen}(1982)}]{Goetz1982}%
  \BibitemOpen
  \bibfield  {author} {\bibinfo {author} {\bibfnamefont {M.}~\bibnamefont
  {Goetz}}\ and\ \bibinfo {author} {\bibfnamefont {J.}~\bibnamefont {Cowen}},\
  }\bibfield  {title} {\enquote {\bibinfo {title} {The thermal conductivity of
  silver iodide},}\ }\href@noop {} {\bibfield  {journal} {\bibinfo  {journal}
  {Solid State Communications}\ }\textbf {\bibinfo {volume} {41}},\ \bibinfo
  {pages} {293--295} (\bibinfo {year} {1982})}\BibitemShut {NoStop}%
\bibitem [{\citenamefont {Malica}\ and\ \citenamefont
  {Dal~Corso}(2019)}]{Malica2019}%
  \BibitemOpen
  \bibfield  {author} {\bibinfo {author} {\bibfnamefont {C.}~\bibnamefont
  {Malica}}\ and\ \bibinfo {author} {\bibfnamefont {A.}~\bibnamefont
  {Dal~Corso}},\ }\bibfield  {title} {\enquote {\bibinfo {title}
  {Temperature-dependent atomic b factor: an ab initio calculation},}\ }\href
  {\doibase https://doi.org/10.1107/S205327331900514X} {\bibfield  {journal}
  {\bibinfo  {journal} {Acta Crystallographica Section A}\ }\textbf {\bibinfo
  {volume} {75}},\ \bibinfo {pages} {624--632} (\bibinfo {year}
  {2019})}\BibitemShut {NoStop}%
\bibitem [{\citenamefont {Young}\ and\ \citenamefont
  {Alder}(1974)}]{Young1974}%
  \BibitemOpen
  \bibfield  {author} {\bibinfo {author} {\bibfnamefont {D.~A.}\ \bibnamefont
  {Young}}\ and\ \bibinfo {author} {\bibfnamefont {B.~J.}\ \bibnamefont
  {Alder}},\ }\bibfield  {title} {\enquote {\bibinfo {title} {Studies in
  molecular dynamics. xiii. singlet and pair distribution functions for
  hard‐disk and hard‐sphere solids},}\ }\href {\doibase 10.1063/1.1681190}
  {\bibfield  {journal} {\bibinfo  {journal} {The Journal of Chemical Physics}\
  }\textbf {\bibinfo {volume} {60}},\ \bibinfo {pages} {1254--1267} (\bibinfo
  {year} {1974})},\ \Eprint
  {http://arxiv.org/abs/https://doi.org/10.1063/1.1681190}
  {https://doi.org/10.1063/1.1681190} \BibitemShut {NoStop}%
\bibitem [{\citenamefont {Talirz}\ \emph {et~al.}(2020)\citenamefont {Talirz},
  \citenamefont {Kumbhar}, \citenamefont {Passaro}, \citenamefont {Yakutovich},
  \citenamefont {Granata}, \citenamefont {Gargiulo}, \citenamefont {Borelli},
  \citenamefont {Uhrin}, \citenamefont {Huber}, \citenamefont {Zoupanos} \emph
  {et~al.}}]{talirz2020materials}%
  \BibitemOpen
  \bibfield  {author} {\bibinfo {author} {\bibfnamefont {L.}~\bibnamefont
  {Talirz}}, \bibinfo {author} {\bibfnamefont {S.}~\bibnamefont {Kumbhar}},
  \bibinfo {author} {\bibfnamefont {E.}~\bibnamefont {Passaro}}, \bibinfo
  {author} {\bibfnamefont {A.~V.}\ \bibnamefont {Yakutovich}}, \bibinfo
  {author} {\bibfnamefont {V.}~\bibnamefont {Granata}}, \bibinfo {author}
  {\bibfnamefont {F.}~\bibnamefont {Gargiulo}}, \bibinfo {author}
  {\bibfnamefont {M.}~\bibnamefont {Borelli}}, \bibinfo {author} {\bibfnamefont
  {M.}~\bibnamefont {Uhrin}}, \bibinfo {author} {\bibfnamefont {S.~P.}\
  \bibnamefont {Huber}}, \bibinfo {author} {\bibfnamefont {S.}~\bibnamefont
  {Zoupanos}},  \emph {et~al.},\ }\bibfield  {title} {\enquote {\bibinfo
  {title} {Materials cloud, a platform for open computational science},}\
  }\href {https://dx.doi.org/10.1038/s41597-020-00637-5} {\bibfield  {journal}
  {\bibinfo  {journal} {Scientific data}\ }\textbf {\bibinfo {volume} {7}},\
  \bibinfo {pages} {1--12} (\bibinfo {year} {2020})}\BibitemShut {NoStop}%
\bibitem [{\citenamefont {Galamba}, \citenamefont {{Nieto de Castro}},\ and\
  \citenamefont {Ely}(2007)}]{Galamba2007}%
  \BibitemOpen
  \bibfield  {author} {\bibinfo {author} {\bibfnamefont {N.}~\bibnamefont
  {Galamba}}, \bibinfo {author} {\bibfnamefont {C.~a.}\ \bibnamefont {{Nieto de
  Castro}}}, \ and\ \bibinfo {author} {\bibfnamefont {J.~F.}\ \bibnamefont
  {Ely}},\ }\bibfield  {title} {\enquote {\bibinfo {title} {{Equilibrium and
  nonequilibrium molecular dynamics simulations of the thermal conductivity of
  molten alkali halides.}}}\ }\href {\doibase 10.1063/1.2734965} {\bibfield
  {journal} {\bibinfo  {journal} {J. Chem. Phys.}\ }\textbf {\bibinfo {volume}
  {126}},\ \bibinfo {pages} {204511} (\bibinfo {year} {2007})}\BibitemShut
  {NoStop}%
\bibitem [{\citenamefont {Bertossa}\ \emph {et~al.}(2019)\citenamefont
  {Bertossa}, \citenamefont {Grasselli}, \citenamefont {Ercole},\ and\
  \citenamefont {Baroni}}]{Bertossa2019}%
  \BibitemOpen
  \bibfield  {author} {\bibinfo {author} {\bibfnamefont {R.}~\bibnamefont
  {Bertossa}}, \bibinfo {author} {\bibfnamefont {F.}~\bibnamefont {Grasselli}},
  \bibinfo {author} {\bibfnamefont {L.}~\bibnamefont {Ercole}}, \ and\ \bibinfo
  {author} {\bibfnamefont {S.}~\bibnamefont {Baroni}},\ }\bibfield  {title}
  {\enquote {\bibinfo {title} {Theory and numerical simulation of heat
  transport in multicomponent systems},}\ }\href {\doibase
  10.1103/PhysRevLett.122.255901} {\bibfield  {journal} {\bibinfo  {journal}
  {Phys. Rev. Lett.}\ }\textbf {\bibinfo {volume} {122}},\ \bibinfo {pages}
  {255901} (\bibinfo {year} {2019})}\BibitemShut {NoStop}%
\bibitem [{\citenamefont {Baroni}\ \emph {et~al.}(2018)\citenamefont {Baroni},
  \citenamefont {Bertossa}, \citenamefont {Ercole}, \citenamefont {Grasselli},\
  and\ \citenamefont {Marcolongo}}]{baroni2018}%
  \BibitemOpen
  \bibfield  {author} {\bibinfo {author} {\bibfnamefont {S.}~\bibnamefont
  {Baroni}}, \bibinfo {author} {\bibfnamefont {R.}~\bibnamefont {Bertossa}},
  \bibinfo {author} {\bibfnamefont {L.}~\bibnamefont {Ercole}}, \bibinfo
  {author} {\bibfnamefont {F.}~\bibnamefont {Grasselli}}, \ and\ \bibinfo
  {author} {\bibfnamefont {A.}~\bibnamefont {Marcolongo}},\ }\enquote {\bibinfo
  {title} {Heat {T}ransport in {I}nsulators from {A}b {I}nitio {G}reen-{K}ubo
  theory},}\ in\ \href {\doibase 10.1007/978-3-319-50257-1_12-1} {\emph
  {\bibinfo {booktitle} {Handbook of Materials Modeling: Applications: Current
  and Emerging Materials}}},\ \bibinfo {editor} {edited by\ \bibinfo {editor}
  {\bibfnamefont {W.}~\bibnamefont {Andreoni}}\ and\ \bibinfo {editor}
  {\bibfnamefont {S.}~\bibnamefont {Yip}}}\ (\bibinfo  {publisher} {Springer
  International Publishing},\ \bibinfo {address} {Cham},\ \bibinfo {year}
  {2018})\ pp.\ \bibinfo {pages} {1--36},\ \bibinfo {edition} {2nd}\ ed.,\
  \Eprint {http://arxiv.org/abs/1802.08006} {arXiv:1802.08006
  [cond-mat.stat-mech]} \BibitemShut {NoStop}%
\bibitem [{\citenamefont {Bonella}, \citenamefont {Ferrario},\ and\
  \citenamefont {Ciccotti}(2017)}]{Bonella2017}%
  \BibitemOpen
  \bibfield  {author} {\bibinfo {author} {\bibfnamefont {S.}~\bibnamefont
  {Bonella}}, \bibinfo {author} {\bibfnamefont {M.}~\bibnamefont {Ferrario}}, \
  and\ \bibinfo {author} {\bibfnamefont {G.}~\bibnamefont {Ciccotti}},\
  }\bibfield  {title} {\enquote {\bibinfo {title} {Thermal diffusion in binary
  mixtures: Transient behavior and transport coefficients from equilibrium and
  nonequilibrium molecular dynamics},}\ }\href {\doibase
  10.1021/acs.langmuir.7b02565} {\bibfield  {journal} {\bibinfo  {journal}
  {Langmuir}\ }\textbf {\bibinfo {volume} {33}},\ \bibinfo {pages}
  {11281--11290} (\bibinfo {year} {2017})}\BibitemShut {NoStop}%
\bibitem [{\citenamefont {Grasselli}\ and\ \citenamefont
  {Baroni}(2021)}]{Grasselli2021}%
  \BibitemOpen
  \bibfield  {author} {\bibinfo {author} {\bibfnamefont {F.}~\bibnamefont
  {Grasselli}}\ and\ \bibinfo {author} {\bibfnamefont {S.}~\bibnamefont
  {Baroni}},\ }\bibfield  {title} {\enquote {\bibinfo {title} {Invariance
  principles in the theory and computation of transport coefficients},}\ }\href
  {\doibase 10.1140/epjb/s10051-021-00152-5} {\bibfield  {journal} {\bibinfo
  {journal} {The European Physical Journal B}\ }\textbf {\bibinfo {volume}
  {94}},\ \bibinfo {pages} {160} (\bibinfo {year} {2021})}\BibitemShut
  {NoStop}%
\bibitem [{\citenamefont {Hirel}(2015)}]{atomsk}%
  \BibitemOpen
  \bibfield  {author} {\bibinfo {author} {\bibfnamefont {P.}~\bibnamefont
  {Hirel}},\ }\bibfield  {title} {\enquote {\bibinfo {title} {Atomsk: A tool
  for manipulating and converting atomic data files},}\ }\href@noop {}
  {\bibfield  {journal} {\bibinfo  {journal} {Computer Physics Communications}\
  }\textbf {\bibinfo {volume} {197}},\ \bibinfo {pages} {212--219} (\bibinfo
  {year} {2015})}\BibitemShut {NoStop}%
\bibitem [{\citenamefont {Puligheddu}\ and\ \citenamefont
  {Galli}(2020)}]{Puligheddu2020}%
  \BibitemOpen
  \bibfield  {author} {\bibinfo {author} {\bibfnamefont {M.}~\bibnamefont
  {Puligheddu}}\ and\ \bibinfo {author} {\bibfnamefont {G.}~\bibnamefont
  {Galli}},\ }\bibfield  {title} {\enquote {\bibinfo {title} {Atomistic
  simulations of the thermal conductivity of liquids},}\ }\href@noop {}
  {\bibfield  {journal} {\bibinfo  {journal} {Physical Review Materials}\
  }\textbf {\bibinfo {volume} {4}},\ \bibinfo {pages} {053801} (\bibinfo {year}
  {2020})}\BibitemShut {NoStop}%
\bibitem [{\citenamefont {Bertossa}(2022)}]{analisi}%
  \BibitemOpen
  \bibfield  {author} {\bibinfo {author} {\bibfnamefont {R.}~\bibnamefont
  {Bertossa}},\ }\href@noop {} {\enquote {\bibinfo {title} {\textsc{analisi}:
  your swiss army knife of molecular dynamics analysis},}\ }\bibinfo
  {howpublished} {\url{https://github.com/rikigigi/analisi}} (\bibinfo {year}
  {2017--2022})\BibitemShut {NoStop}%
\bibitem [{\citenamefont {Brehm}\ and\ \citenamefont
  {Kirchner}(2011)}]{travis1}%
  \BibitemOpen
  \bibfield  {author} {\bibinfo {author} {\bibfnamefont {M.}~\bibnamefont
  {Brehm}}\ and\ \bibinfo {author} {\bibfnamefont {B.}~\bibnamefont
  {Kirchner}},\ }\bibfield  {title} {\enquote {\bibinfo {title} {Travis-a free
  analyzer and visualizer for monte carlo and molecular dynamics
  trajectories},}\ }\href {\doibase https://doi.org/10.1021/ci200217w}
  {\bibfield  {journal} {\bibinfo  {journal} {J. Chem. Inf. Model.}\ }\textbf
  {\bibinfo {volume} {51}},\ \bibinfo {pages} {2007–2023} (\bibinfo {year}
  {2011})}\BibitemShut {NoStop}%
\bibitem [{\citenamefont {Brehm}\ \emph {et~al.}(2020)\citenamefont {Brehm},
  \citenamefont {Thomas}, \citenamefont {Gehrke},\ and\ \citenamefont
  {Kirchner}}]{travis2}%
  \BibitemOpen
  \bibfield  {author} {\bibinfo {author} {\bibfnamefont {M.}~\bibnamefont
  {Brehm}}, \bibinfo {author} {\bibfnamefont {M.}~\bibnamefont {Thomas}},
  \bibinfo {author} {\bibfnamefont {S.}~\bibnamefont {Gehrke}}, \ and\ \bibinfo
  {author} {\bibfnamefont {B.}~\bibnamefont {Kirchner}},\ }\bibfield  {title}
  {\enquote {\bibinfo {title} {Travis—a free analyzer for trajectories from
  molecular simulation},}\ }\href {\doibase https://doi.org/10.1063/5.0005078}
  {\bibfield  {journal} {\bibinfo  {journal} {The Journal of chemical physics}\
  }\textbf {\bibinfo {volume} {152}},\ \bibinfo {pages} {164105} (\bibinfo
  {year} {2020})}\BibitemShut {NoStop}%
\bibitem [{\citenamefont {Fan}\ \emph {et~al.}(2015)\citenamefont {Fan},
  \citenamefont {Pereira}, \citenamefont {Wang}, \citenamefont {Zheng},
  \citenamefont {Donadio},\ and\ \citenamefont {Harju}}]{Fan2015}%
  \BibitemOpen
  \bibfield  {author} {\bibinfo {author} {\bibfnamefont {Z.}~\bibnamefont
  {Fan}}, \bibinfo {author} {\bibfnamefont {L.~F.~C.}\ \bibnamefont {Pereira}},
  \bibinfo {author} {\bibfnamefont {H.-Q.}\ \bibnamefont {Wang}}, \bibinfo
  {author} {\bibfnamefont {J.-C.}\ \bibnamefont {Zheng}}, \bibinfo {author}
  {\bibfnamefont {D.}~\bibnamefont {Donadio}}, \ and\ \bibinfo {author}
  {\bibfnamefont {A.}~\bibnamefont {Harju}},\ }\bibfield  {title} {\enquote
  {\bibinfo {title} {Force and heat current formulas for many-body potentials
  in molecular dynamics simulations with applications to thermal conductivity
  calculations},}\ }\href@noop {} {\bibfield  {journal} {\bibinfo  {journal}
  {Physical Review B}\ }\textbf {\bibinfo {volume} {92}},\ \bibinfo {pages}
  {094301} (\bibinfo {year} {2015})}\BibitemShut {NoStop}%
\bibitem [{\citenamefont {Ercole}, \citenamefont {Marcolongo},\ and\
  \citenamefont {Baroni}(2017)}]{Ercole2017}%
  \BibitemOpen
  \bibfield  {author} {\bibinfo {author} {\bibfnamefont {L.}~\bibnamefont
  {Ercole}}, \bibinfo {author} {\bibfnamefont {A.}~\bibnamefont {Marcolongo}},
  \ and\ \bibinfo {author} {\bibfnamefont {S.}~\bibnamefont {Baroni}},\
  }\bibfield  {title} {\enquote {\bibinfo {title} {{Accurate thermal
  conductivities from optimally short molecular dynamics simulations}},}\
  }\href {\doibase 10.1038/s41598-017-15843-2} {\bibfield  {journal} {\bibinfo
  {journal} {Sci. Rep.}\ }\textbf {\bibinfo {volume} {7}},\ \bibinfo {pages}
  {15835} (\bibinfo {year} {2017})},\ \Eprint {http://arxiv.org/abs/1706.01381}
  {arXiv:1706.01381} \BibitemShut {NoStop}%
\bibitem [{\citenamefont {Boone}, \citenamefont {Babaei},\ and\ \citenamefont
  {Wilmer}(2019)}]{Boone2019}%
  \BibitemOpen
  \bibfield  {author} {\bibinfo {author} {\bibfnamefont {P.}~\bibnamefont
  {Boone}}, \bibinfo {author} {\bibfnamefont {H.}~\bibnamefont {Babaei}}, \
  and\ \bibinfo {author} {\bibfnamefont {C.~E.}\ \bibnamefont {Wilmer}},\
  }\bibfield  {title} {\enquote {\bibinfo {title} {Heat flux for many-body
  interactions: corrections to lammps},}\ }\href@noop {} {\bibfield  {journal}
  {\bibinfo  {journal} {Journal of chemical theory and computation}\ }\textbf
  {\bibinfo {volume} {15}},\ \bibinfo {pages} {5579--5587} (\bibinfo {year}
  {2019})}\BibitemShut {NoStop}%
\bibitem [{\citenamefont {Surblys}\ \emph {et~al.}(2019)\citenamefont
  {Surblys}, \citenamefont {Matsubara}, \citenamefont {Kikugawa},\ and\
  \citenamefont {Ohara}}]{Surblys2019}%
  \BibitemOpen
  \bibfield  {author} {\bibinfo {author} {\bibfnamefont {D.}~\bibnamefont
  {Surblys}}, \bibinfo {author} {\bibfnamefont {H.}~\bibnamefont {Matsubara}},
  \bibinfo {author} {\bibfnamefont {G.}~\bibnamefont {Kikugawa}}, \ and\
  \bibinfo {author} {\bibfnamefont {T.}~\bibnamefont {Ohara}},\ }\bibfield
  {title} {\enquote {\bibinfo {title} {Application of atomic stress to compute
  heat flux via molecular dynamics for systems with many-body interactions},}\
  }\href@noop {} {\bibfield  {journal} {\bibinfo  {journal} {Physical Review
  E}\ }\textbf {\bibinfo {volume} {99}},\ \bibinfo {pages} {051301} (\bibinfo
  {year} {2019})}\BibitemShut {NoStop}%
\bibitem [{\citenamefont {Simmons}, \citenamefont {Wang}\ \emph
  {et~al.}(1971)\citenamefont {Simmons}, \citenamefont {Wang} \emph
  {et~al.}}]{Simmons1971}%
  \BibitemOpen
  \bibfield  {author} {\bibinfo {author} {\bibfnamefont {G.}~\bibnamefont
  {Simmons}}, \bibinfo {author} {\bibfnamefont {H.}~\bibnamefont {Wang}},
  \emph {et~al.},\ }\href@noop {} {\emph {\bibinfo {title} {Single crystal
  elastic constants and calculated aggregate properties}}}\ (\bibinfo
  {publisher} {Mass., MIt Press},\ \bibinfo {year} {1971})\BibitemShut
  {NoStop}%
\end{thebibliography}
\end{document}